\documentclass[preprint,aps,prd,showpacs,preprintnumbers,amsmath,amssymb,amsfonts,nofootinbib,a4paper,12pt]{revtex4-1}

\usepackage{graphicx}
\usepackage[usenames]{color}
\usepackage[usenames,dvipsnames]{xcolor}

\usepackage{hyperref}

\numberwithin{equation}{section}

\def\stamp{--- {\bf \today} --- {\bf \jobname.tex}}

\def\tr{\textrm{tr}}

\def\cT{\mathcal{T}}
\def\cA{\mathcal{A}}
\def\cM{\mathcal{M}}
\def\cN{\mathcal{N}}
\def\cP{\mathcal{P}}

\def\cS{\mathcal{S}}
\def\cZ{\mathcal{Z}}

\def\IC{\mathbb{C}}
\def\ZZ{\mathbb{Z}}

\def\abb{\left[\bar a \atop \bar b \right]}

\def\tr{\textrm{tr}}

\def\tr{\textrm{tr}}

\def\cF{\mathcal{F}}
\def\nn{\nonumber}
\def\sg(#1){\textrm{sign}(#1)}

\def\cP{\mathcal{P}}

\def\cN{\mathcal{N}}
\def\cA{\mathcal{A}}
\def\cM{\mathcal{M}}
\def\cW{\mathcal{W}}
\def\cQ{\mathcal{Q}}

\def\ZZ{\mathbb{Z}}
\def\ie{\emph{i.e.}}

\newcommand{\abd}[2]{\left[#1  \atop #2 \right]}

\def\an[#1,#2]{\left\langle#1\,#2\right\rangle}
\def\aq[#1,#2,#3]{\left\langle#1|#2|#3\right]}
\def\qa[#1,#2,#3]{\left[#1|#2|#3\right\rangle}
\def\sq[#1,#2]{\left[#1\,#2\right]}
\def\spa#1.#2{\left\langle#1\,#2\right\rangle}
\def\spab#1.#2.#3{\left\langle#1|#2|#3\right]}
\def\spb#1.#2{\left[#1\,#2\right]}
\def\lor#1.#2{\left(#1\,#2\right)}

\begin{document}
\preprint{IHES/P/12/21, IPHT-t12/057}
\title{One-loop four-graviton amplitudes in $\mathcal N=4$ supergravity models}
\author{{\bf Piotr Tourkine}${}^a$ and {\bf Pierre Vanhove}$^{a,b}$}
\affiliation{ ${}^a$ Institut de Physique Th{\'e}orique, CEA/Saclay, F-91191 Gif-sur-Yvette,
 France\\
${}^b$ IHES,  Le  Bois-Marie, 35  route de  Chartres,
F-91440 Bures-sur-Yvette, France\\
{\tt email: piotr.tourkine,pierre.vanhove@cea.fr}}

\begin{abstract}
We evaluate in great detail one-loop four-graviton field theory amplitudes in
pure  $\cN=4$ $D=4$ supergravity. The expressions are obtained by
taking the field theory limits of $(4,0)$ and $(2,2)$ space-time
supersymmetric string theory models. 
For each model we extract the contributions of the spin-1
and spin-2 $\cN=4$ supermultiplets running in the loop.
We show that all of those constructions lead to the same four-dimensional result
for the four-graviton amplitudes in pure supergravity even though they come from
different string theory models. 
\end{abstract}

\maketitle

\section{Introduction}
\label{sec:Introduction}


The role of supersymmetry in perturbative supergravity still leaves
room for surprises. The construction of candidate counter-terms for
ultraviolet (UV) divergences in extended four-dimensional supergravity theories
does not forbid some particular amplitudes to have an  improved  UV
behaviour. For instance, the four-graviton three-loop amplitude in $\cN=4$
supergravity turns out to be UV finite~\cite{Bern:2012cd,Tourkine:2012ip},
despite the construction of a candidate counter-term~\cite{Bossard:2011tq}. (Some early discussion of the
three-loop divergence  in $\cN=4$ has appeared in~\cite{Deser:1977nt}, and
recent alternative arguments have been given in~\cite{Kallosh:2012ei}.)

The UV behaviour of extended supergravity theories is constrained in string
theory by non-renormalisation theorems that give rise in the field theory limit
to supersymmetric protection for potential counter-terms. In maximal 
supergravity, the absence of divergences until
six loops in four dimensions~\cite{Bern:2007hh,Bern:2008pv,Bern:2009kd}
is indeed a consequence of the supersymmetric protection for $\frac12$-,
$\frac14$- and $\frac18$-BPS operators in
string~\cite{Green:2006gt,Green:2010sp} or field
theory~\cite{Bossard:2009sy,Bossard:2010bd}. In half-maximal 
supergravity, it was
shown recently~\cite{Tourkine:2012ip} that the absence of three-loop
divergence in the four-graviton amplitude in four dimensions is a consequence of
the protection of the $\frac12$-BPS $R^4$ coupling from perturbative quantum
corrections beyond one loop in heterotic
models. We refer to~\cite{Tseytlin:1995bi,Bachas:1997mc,D'Hoker:2005jc} for a
discussion of the non-renormalisation theorems in heterotic string.

Maximal  supergravity is unique in any dimension, and corresponds to the
massless sector of 
type II string theory compactified on a torus. Duality symmetries relate
different phases of the theory and strongly constrain its UV
behaviour~\cite{Green:2010sp,Elvang:2010jv,Elvang:2010kc,Beisert:2010jx,
Elvang:2010xn,Bossard:2010bd}.

On the contrary, half-maximal supergravity (coupled to vector multiplets) is
not unique and can be obtained in the low-energy limit of
$(4,0)$ string theory models---with all the space-time supersymmetries coming
from the
world-sheet left-moving sector---or $(2,2)$ string theory models---with
the space-time supersymmetries originating both from the world-sheet left-moving and right-moving
sectors. The two constructions  give rise to different low-energy supergravity
theories with a different identification of the moduli.

In this work we analyze the properties of the four-graviton amplitude
at one loop in pure $\cN=4$ supergravity in four dimensions. We compute the genus one
string theory amplitude in different models and extract its field theory
limit. 
This method has been pioneered by \cite{Green:1982sw}. It has then been
developed intensively for gauge theory amplitudes
by~\cite{Bern:1990cu,Bern:1990ux}, and then applied to gravity amplitudes
in~\cite{Bern:1993wt,Dunbar:1994bn}. In this work we will
follow more closely the formulation given in~\cite{BjerrumBohr:2008ji}.

We consider three classes of four-dimensional string
models. The first class, on which was based the analysis
in~\cite{Tourkine:2012ip}, are heterotic string models. They have
$(4,0)$ supersymmetry and $4\leq n_v\leq 22$ vector multiplets. The
models of the second class also carry $(4,0)$ supersymmetry; they are
type~II asymmetric orbifolds. We will study a model with $n_v=0$
(the Dabholkar-Harvey construction, see~\cite{Dabholkar:1998kv}) and a model with $n_v=6$.
 The third class is composed
of type~II symmetric orbifolds with $(2,2)$
supersymmetry. For a given number of vector multiplets, the $(4,0)$
models are related to one another by strong-weak S-duality and related
to $(2,2)$ models by U-duality~\cite{Witten:1995ex,Hull:1994ys}. Several tests of the duality relations
between orbifold models have been given in~\cite{Gregori:1997hi}.

The string theory constructions generically contain matter vector
multiplets. By comparing models with $n_v \neq0$ vector multiplets to a model
where $n_v=0$, we directly check that one can simply subtract these
contributions and
extract the pure $\cN=4$ supergravity contributions in four
dimensions.

We shall show that the four-graviton amplitudes extracted from the
$(4,0)$ string models match that obtained in~\cite{Dunbar:1994bn,Dunbar:1995ed,Dunbar:2010fy,Dunbar:2011xw,Dunbar:2011dw,Dunbar:2012aj,Bern:2011rj}. We however note that 
all of those constructions are based on a $(4,0)$ construction, while our
analysis covers both the $(4,0)$ and a $(2,2)$ models. The
four-graviton amplitudes are expressed
in a supersymmetric decomposition into $\cN=4\,s$ spin-$s$ supermultiplets with
$s=1,\frac32,2$, as in
\cite{Dunbar:1994bn,Dunbar:1995ed,Dunbar:2010fy,Dunbar:2011xw,Dunbar:2011dw,
Dunbar:2012aj,Bern:2011rj}. The $\cN=8$ and $\cN=6$ supermultiplets have the
same integrands in all the models, while the contributions of the
$\cN=4$ multiplets have different integrands. Despite the absence of
obvious relation between the integrands of the two
models, the amplitudes turn
out to be equal after integration in all the string theory models.
In a nutshell, we find that the four-graviton one-loop field theory amplitudes
in the $(2,2)$ construction are identical to the $(4,0)$ ones.

The paper is organized as follows. For each model we evaluate the one-loop
four-graviton string theory amplitudes in section~\ref{sec:1loop}. In
section~\ref{sec:comparing} we compare the expressions that we obtained and
check that they are compatible with our expectations from string dualities. We
then extract and evaluate the field theory limit in the regime $\alpha'\to0$ of
those string amplitudes in section~\ref{sec:qftlimit-one-loop}. This gives us
the field theory four-graviton one-loop amplitudes for pure $\cN=4$
supergravity.
Section~\ref{sec:conclusion} contains our conclusions. Finally, 
Appendices~\ref{sec:chiralblocks} and~\ref{sec:chblocks} contain
details about our conventions and the properties of the conformal field
theory (CFT) building
blocks of our string theory models.

\section{One-loop String theory amplitudes in  $(4,0)$ and $(2,2)$ models}
\label{sec:1loop}
\label{sec:one-loop-string}

In this section, we compute the one-loop four-graviton amplitudes in four-dimensional $\cN=4$ $(4,0)$ and $(2,2)$ string theory models. Their massless
spectrum contains an $\cN=4$ supergravity multiplet coupled to $n_v$
$\cN=4$
vector multiplets. Since the heterotic string is made of the tensor product of a
left-moving superstring by a right-moving bosonic string, it only gives rise to
$(4,0)$ models. However, type~II compactifications  provide the
freedom to build  $(4,0)$ and $(2,2)$ models~\cite{Ferrara:1989nm}.

\subsection{Heterotic CHL models}
\label{sec:one-loop-CHL}

We evaluate the one-loop four-graviton amplitudes in
heterotic string CHL models in four dimensions~\cite{Chaudhuri:1995bf,Schwarz:1995bj,Aspinwall:1995fw}. Their low-energy limits are
$(4,0)$ supergravity models with $4\leq n_v\leq 22$ vector supermultiplets matter
fields. We first comment on the moduli space of the model, then write
the string theory one-loop amplitude and finally compute the CHL
partition function. This allows us to extract the massless states contribution
 to the integrand of the field theory limit.

These models have the following moduli space:
\begin{equation}\label{e:modspace}
  \Gamma\backslash SU(1,1)/U(1) \times    SO(6,n_v;\ZZ) \backslash
  SO(6,n_v)/SO(6)\times SO(n_v)\,,
\end{equation}
where $n_v$ is the number of vector multiplets, and
 $\Gamma$ is a discrete subgroup of $SL(2,\ZZ)$.
For instance, 
$\Gamma=SL(2,\ZZ)$ for $n_v=22$ and  $\Gamma=\Gamma_1(N)$ for the
$\ZZ_N$ CHL $(4,0)$ orbifold. (We refer to 
Appendix~\ref{sec:congruence} for a definition of the congruence
subgroups of $SL(2,\ZZ)$.) The scalar manifold $SU(1,1)/U(1)$ is
parametrized by the axion-dilaton in the $\cN=4$ gravity
supermultiplet.

 The generic structure of the amplitude has been described in~\cite{Tourkine:2012ip}.  We will
use the same notations and conventions.  The four-graviton amplitude takes the
following form\footnote{The $t_8$ tensor defined
  in~\cite[appendix~9.A]{Green:1987mn} is given by $t_8F^4=4\tr(F^{(1)}F^{(2)}F^{(3)}F^{(4)})-
  \tr(F^{(1)}F^{(2)})\tr(F^{(3)}F^{(4)})+perms(2,3,4)$, where the traces are
  taken over the Lorentz indices. Setting the coupling constant to
  one, $t_8F^4=st A^{tree}(1,2,3,4)$ where $A^{tree}(1,2,3,4)$ is the color stripped
ordered tree amplitude between four gluons.}
\begin{equation}
  \label{e:4gravHet}
 \mathcal M^{(n_v)}_{(4,0)het} =\cN\,\left(\pi\over2\right)^4t_8 F^4 \, \int_{\cF}
{d^2\tau\over\tau_2^{D-6\over2}}\, \int_{\cT}
  \prod_{1\leq i<j\leq 4} {d^2\nu_i\over\tau_2}\,
  e^{\mathcal Q}\,
  \cZ^{(n_v)}_{(4,0)het}\bar\cW^B \,,
\end{equation}
where $D=10-d$,  and $\cN$ is the normalization constant  of the amplitude.
The domains of integration are $\cF=\{\tau=\tau_1+i\tau_2;
|\tau_1|\leq \frac12, |\tau|^2\geq1, \tau_2>0\}$ and
$\cT:=\{\nu=\nu_1+i\nu_2; |\nu_1|\leq\frac12, 0\leq \nu_2\leq\tau_2\}$. Then,
\begin{equation}\label{e:Wgenus1}
\bar   \cW^B:=\, {\langle \prod_{j=1}^4
\tilde\epsilon^j \cdot \bar\partial X(\nu_j)
    e^{i k_j\cdot X(\nu_j)} \rangle\over (2 \alpha')^4 \langle  \prod_{j=1}^4 
    e^{i k_j\cdot X(\nu_j)}  \rangle}
\end{equation}
is the kinematical factor coming from the Wick contractions of  the bosonic
vertex operators  and the plane-wave part is given by
$\langle  \prod_{j=1}^4     e^{i k_j\cdot X(\nu_j)}  \rangle= \exp(\cQ)$
with 
\begin{equation}
  \label{e:QstringDef}
\mathcal Q=  \sum_{1\leq i<j\leq4}2\alpha' k_i\cdot k_j \cP(\nu_{ij})\,,
\end{equation}
where we have made use of the notation $\nu_{ij}:=\nu_i-\nu_j$.
Using the result of~\cite{Sakai:1986bi} with our
normalizations we explicitly write
\begin{equation}
  \label{e:thatone}
\bar\cW^B= \prod_{r=1}^4 \tilde\epsilon_r\cdot \bar\cQ_r+
{1\over2\alpha'}
  \,(\tilde\epsilon_1\cdot \bar\cQ_1\,
  \tilde\epsilon_2\cdot \bar\cQ_2\,\tilde\epsilon_3\cdot \tilde\epsilon_4
  \bar\cT(\nu_{34})+perms)+{1\over4{\alpha'}^2}\,
  (\tilde\epsilon_1\cdot\tilde\epsilon_2\, \tilde\epsilon_3\cdot\tilde\epsilon_4
\, \bar\cT(\nu_{12})\bar\cT(\nu_{34})+perms)\,,
\end{equation}
where we have introduced
\begin{equation}\label{e:QT}
  \cQ_I^\mu:= \sum_{r=1}^4 \, k^{\mu}_r\, \partial
  \cP(\nu_{Ir}|\tau);\qquad 
\cT(\nu):= \partial_\nu^2 \cP(\nu|\tau)\,,
\end{equation}
with $\cP(z)$  the genus one bosonic propagator. We refer to
Appendix~\ref{sec:green-functions} for definitions and conventions.

The CHL models studied in this work are asymmetric
$\ZZ_N$ orbifolds of the bosonic sector (in our  case the
right-moving sector) of the heterotic string compactified on
$T^5\times S^1$. Geometrically, the orbifold rotates $N$ groups of $\ell$
bosonic
fields $\bar X^a$ belonging either to the internal $T^{16}$ or to the
$T^5$ and acts as an order $N$ shift on the $S^1$. More precisely, if we take a
boson
$\bar X^a$ of the $(p+1)$-th group ($p=0,\ldots,N-1$) of $\ell$ bosons, we have
$a\in\{p\ell,p\ell+1,\ldots,p\ell+(\ell-1)\}$ and for twists $g/2,h/2\in
\{0,1/N,\ldots,(N-1)/N\}$ we get 
\begin{eqnarray}
\bar X^a(z+\tau)&=& e^{ i \pi g p/N} \bar X^a(z)\,,\nn\\
\bar X^a(z+1) &=& e^{ i \pi h p/N} \bar X^a(z)\,.
\end{eqnarray}
We will consider models with $(N,n_v,\ell)\in\{(1,22,16), (2,14,8), (3,10,6),
(5,6,4),\break (7,4,3)\}$. It is in principle possible to build models with
$(N,n_v,\ell)=(11,2,2)$ and $(N,n_v,\ell)=(23,0,1)$ and thus decouple totally
the matter fields, but it is then required to compactify the theory on
a seven- and eight-dimensional torus respectively. We will not comment about it
further, since we have anyway a type II superstring compactification with
$(4,0)$ supersymmetry
that already has $n_v=0$ that we discuss in
section~\ref{sec:models-with-zero}. This issue could have been important, but
it appears that at one loop in the field theory limit there are no problem to decouple the
vector multiplets to obtain pure $\cN=4$ supergravity. The partition function
of the right-moving CFT is given by
\begin{equation}
\cZ_{(4,0)het}^{(n_v)}(\tau)=\frac{1}{|G|} \sum_{(g,h)}  \cZ_{(4,0)het}^{h,g}(\tau)\,,
\end{equation}
where $|G|$ is the order of the orbifold group \ie~$|G|=N$. The
twisted conformal blocks
$\cZ^{g,h}_{het}$ are a product of the oscillator and zero mode part
\begin{equation}
\label{e:PFCHL}
 \cZ^{h,g}_{(4,0)het}  = \cZ^{h,g}_{osc}\times \cZ^{h,g}_{latt}\,.
\end{equation}
In the field theory limit only the massless states from the $h=0$ sector will
contribute and we are left with:
\begin{equation}
  \cZ_{(4,0)het}^{(n_v)}(\tau)\to
  {1 \over N} \,\cZ_{(4,0)het}^{0,0}(\tau)
+{1\over N} \sum_{\{g\}}\cZ_{(4,0)het}^{0,g}(\tau)\,.
\label{e:czhetlim}
\end{equation}

The untwisted partition function ($g=h=0$)  with generic diagonal
Wilson lines $A$, as required by modular invariance, is 
\begin{equation}
  \cZ_{(4,0)het}^{0,0}(\tau):= 
{\Gamma_{(6,24)}(G,A)\over\bar\eta^{24}(\bar\tau)}\,,
\end{equation}
where $\Gamma_{(6,24)}(G,A)$ is the lattice sum for the Narain lattice
$ \Gamma^{(5,5)}\oplus \Gamma^{(1,1)} \oplus  \Gamma_{E_8\times E_8}$
with Wilson lines~\cite{Narain:1986am}. It drops out in the field theory limit
where the radii of compactification
$R\sim\sqrt{\alpha'}$ are sent to zero and we are left with the part coming from
the
oscillators
\begin{equation}
     \cZ_{(4,0)het}^{0,0}(\tau)\to {1\over \bar\eta^{24}(\bar\tau)}\,.
\end{equation}
At a generic point in the moduli space, the $480$
gauge bosons of the adjoint representation of $E_8 \times E_8$ get masses due to Wilson
lines, and only the $\ell$ gauge bosons of the
$U(1)^\ell$ group left invariant by the orbifold
action~\cite{Jatkar:2005bh,Dabholkar:2006bj} stay in the matter massless
spectrum.

The oscillator part is computed to be
\begin{equation}
 \cZ^{h,g}_{osc}=\sum_{\{g,h\}}\,\prod_{p=0}^{N-1}\left(\cZ^{h\times      
p,g\times p}_{X}\right)^{ \ell }\,,
\end{equation}
where the twisted bosonic chiral blocks $\cZ_X^{h,g}$ are given in
Appendix~\ref{sec:chiralblocks}. For $h=0$, $\cZ^{0,g}_{osc}$ is independent of
$g$ when $N$ is prime and it can be computed explicitly. It is the inverse of
the unique cusp form $f_k(\tau)=(\eta(\tau)\eta(N \tau))^{k+2}$ for
$\Gamma_1 (N)$ of modular weight $\ell=k+2=24/(N+1)$ with $n_v=2\ell-2$ as
determined in~\cite{Jatkar:2005bh,Dabholkar:2006bj}. Then \eqref{e:czhetlim}
writes 
\begin{equation}
\label{e:Z1nv} \cZ_{(4,0)het}\to \frac 1 N \left( {1 \over (\bar \eta(\bar
\tau))^{24}} +  {N-1\over
f_k(\bar \tau)}\right)\,.
\end{equation}

To conclude this section, we write
the part of the integrand of~\eqref{e:4gravHet} that will contribute in the
field theory limit. When $\alpha'\to0$, the region of the fundamental domain of integration
$\cF$ of interest is the large $\tau_2$ region, such that $t=\alpha' \tau_2$
stays constant. Then, the objects that we have introduced admit an expansion in
the
variable $q=e^ {2 i \pi \tau}\to0$. We find
\begin{equation}
\label{e:Z1nvqexp} \cZ_{(4,0)het}\to  {1\over \bar q}+2+n_v +o(\bar q)\,.
\end{equation}
Putting everything together and using the expansions given in~\eqref{e:delQ}, we find that the integrand
in~\eqref{e:4gravHet} is given by 
\begin{equation}
\cZ_{(4,0)het} \cW^B e^\cQ\to e^{\pi \alpha'\tau_2 Q}\left(
 \left(\cW^B e^{    \cQ}\right)|_{\bar q} +(n_v+2)
\left(\cW^Be^{\cQ}\right)|_{\bar q^0} +o(\bar q)\right)\,.
\end{equation}
Order $\bar q$ coefficients are present because of the $1/\bar
q$
chiral tachyonic pole in the non-supersymmetric sector of the theory. Since
the integral over $\tau_1$ of
$\bar q^{-1}\,\left(\cW^Be^{\cQ}\right)|_{\bar q^0}$ vanishes, as a
consequence of  the level matching condition,  we
did not write it.
We introduce $\cA$,   the massless sector contribution to 
the field theory limit of the amplitude at the leading order in
$\alpha'$, for later use in sections~\ref{sec:comparing} and~\ref{sec:qftlimit-one-loop} 
\begin{equation}
 \label{e:amp-het} 
\cA^{(n_v)}_{(4,0)het}=\frac12\,\left(\pi\over2\right)^4\,t_8F^4\,
\left(\bar\cW^B|_{\bar q} (1+\alpha'\delta\cQ) + \bar\cW^B|_{\bar q^0} \cQ|_{\bar q} +(n_v+2)\bar \cW^B|_{\bar q^0}\right)\,,
\end{equation}
where  we have made use  of the notations for the $\bar q$ expansion
\begin{eqnarray}
 \bar \cW^B  &=& \bar\cW^B|_{ q^0}+ \bar q \, \bar\cW^B|_{ q}+o(\bar q^2)\,,\cr
 \cQ &=&-\pi \, \alpha'\tau_2\,Q +\alpha'\delta Q+ q\, \cQ|_{ q}+ \bar q\, \cQ|_{\bar q}+o(|q|^2)\,.
 \end{eqnarray}
%

\subsection{Type~II asymmetric orbifold}
\label{sec:one-loop-asym}

In this section we consider type~II string theory on two different kinds of
asymmetric orbifolds. They lead to $(4,0)$ models with a moduli space given
in~\eqref{e:modspace}, where the axion-dilaton parametrizes the
$SU(1,1)/U(1)$ factor. The first one is a $\ZZ_2$
orbifold with $n_v=6$. The others are the Dabholkar-Harvey
models~\cite{Dabholkar:1998kv,Mizoguchi:2001cp}; they have $n_v=0$
vector multiplet. 

First, we give a general formula for the treatment of those asymmetric
orbifolds.  We then study in detail the partition function of two
particular models and extract the contribution of massless states to
the integrand of the field theory limit of their amplitudes.
A generic expression for the scattering amplitude of four gravitons at
one loop in type IIA and IIB superstring is: 
\begin{eqnarray}
 \label{e:4grav-IIasym} \cM^{(n_v)}_{(4,0)II}&=&\cN\,\int_{\cF}
 {d^2\tau\over\tau_2^{D- 6\over 2}}\int_{\cT}\prod_{1\leq i<j\leq4}
 {d^2\nu_i\over\tau_2}\, e^{\mathcal Q}\, \times\\
\nn&\times& \frac12\sum_{a,b=0,1}(-1)^{a+b+ab} \cZ_{a, b} \cW_{a,b}
\times\\
\nn&\times&\frac1{2|G|}\sum_{    \bar a,\bar b=0,1\atop g,h}(-1)^{\bar a+\bar
    b+\mu \bar a\bar b} (-1)^{C(\bar a,\bar b,g,h)} \,\bar \cZ_{\bar
a, \bar b}^{h,g}
\tilde \cW_{\bar a,\bar b}\,,
\end{eqnarray}
where $\cN$ is the same normalization factor as for the heterotic
string amplitude and $C(\bar a,\bar b,g,h)$ is a model-dependent phase factor
determined by modular invariance and discussed below. %
We have introduced the chiral partition functions in the $(a,b)$-spin
structure
\begin{equation}
 \cZ_{a, b}={\theta\left[a\atop b\right](0|\tau)^4 \over \eta(\tau)^{12}}\,;\qquad
\cZ_{1,1}=0\,.
\label{e:chblkuntw}
\end{equation}
The value of $\mu$ determines the chirality of the  theory: $\mu=0$ for type~IIA and $\mu=1$ for
type~IIB. 
The partition function in a twisted sector $(h,g)$ of
the orbifold is denoted $\bar\cZ^{h,g}_{\bar a,\bar b}$. Notice that the four-dimensional fermions are not twisted, so the vanishing of their partition
function in the $(a,b)=(1,1)$
sector holds for a $(g,h)$-twisted sector: $\bar\cZ^{h,g}_{1,1}=0$.
This is fully consistent with the fact that due to the lack of
fermionic zero modes, this amplitude does not receive any
contributions from the odd/odd, odd/even or even/odd spin structures.
We use the  holomorphic factorization of the $(0,0)$-ghost picture graviton
vector operators as  
\begin{equation}
 V^{(0,0)}=\int d^2z \,  :\epsilon^{(i)}\cdot
 V(z) \, \tilde \epsilon^{(i)}\cdot\bar  V(\bar z)\, e^{ik\cdot X(z,\bar z)}:   
\,,
\end{equation}
with
\begin{equation}
\epsilon^{(i)}\cdot V(z) = \epsilon^{(i)}\cdot\partial X -i
\frac{F^{(i)}_{\mu\nu}}{2} :\psi^\mu
\psi^\nu:;\qquad
\tilde\epsilon^{(i)}\cdot\bar V(\bar z)
=\tilde\epsilon^{(i)}\cdot\bar \partial X +i{\tilde
F^{(i)}_{\mu\nu}\over 2}:\bar\psi^\mu
\bar\psi^\nu:\,,
\end{equation}
where we have introduced the field strengths $F^{(i)}_{\mu\nu}=\epsilon^{(i)}_\mu
k_{i\,\nu}-\epsilon^{(i)}_\nu k_{i\,\mu}$ and $\tilde F^{(i)}_{\mu\nu}=\tilde\epsilon^{(i)}_\mu
k_{i\,\nu}-\tilde\epsilon^{(i)}_\nu k_{i\,\mu}$.

The correlators of the vertex operators in the  $(a,b)$-spin structure are given  by
 $\cW_{a,b}$ and $\bar\cW_{\bar a,\bar b}$ defined by, respectively,
\begin{equation}
  \label{e:W1ab}
 \cW_{a,b}={\langle \prod_{j=1}^4 \epsilon^{(j)}\cdot V(z_j) \,e^{i k_j\cdot X(z_j)} \rangle_{a,b}\over
(2\alpha')^4\langle \prod_{j=1}^4 e^{i k_j\cdot X(z_j)}  \rangle}\,,\qquad
\bar \cW_{\bar a,\bar b}={\langle \prod_{j=1}^4 \tilde \epsilon^{(j)}\cdot \bar
V(\bar z_j) \,e^{i k_j\cdot X(z_j)} \rangle_{\bar a,\bar b}\over(2\alpha')^4
\langle \prod_{j=1}^4 e^{i k_j\cdot X(\bar z_j)}  \rangle}\,.
\end{equation}

We decompose the $\cW_{a,b}$ into one part that depends on the spin
structure $(a,b)$, denoted $\cW^F_{a,b}$,  and another  independent of
the  spin structure $\cW^B$:
\begin{equation}
\cW_{a,b}= \cW^F_{a,b}+\cW^B\,,
\label{e:Wabdef}
\end{equation}
this last term being identical to the one given
in~\eqref{e:Wgenus1}. 
The spin structure-dependent part is given by the following fermionic Wick's
contractions:
\begin{equation}
\cW_{a,b}^F= \cS_{4;a,b}+ \cS_{2;a,b}\,,
\label{e:WFabdef}
\end{equation}
where $\cS_{n;a,b}$ arise from Wick contracting $n$ pairs of
world-sheet fermions. Note that the contractions involving three pairs of fermion
turn out to vanish in all the type II models by symmetry.
We introduce the notation $\sum_{\{(i,\cdots),(j,\cdots)\}=\{1,2,3,4\}}
\cdots$ for the sum over the ordered partitions of $\{1,2,3,4\}$ into two
sets where the partitions $\{(1,2,3),1\}$ and $\{(1,3,2),1\}$
are considered to be independent. In that manner, the two terms
in~\eqref{e:WFabdef} can be written explicitly:
\begin{eqnarray}
 \cS_{4;a,b} &=&{1\over2^{10}} \sum_{\{(i,j),(k,l)\}=\{1,2,3,4\}}
S_{a,b} (z_{ij}) S_{a,b} (z_{ji})
S_{a,b} (z_{kl}) S_{a,b} (z_{lk}) \,\tr(F^{(i)}F^{(j)})
\,\tr(F^{(k)}F^{(l)})\nn\\
&-&{1\over 2^8}\sum_{\{(i,j,k,l)\}=\{1,2,3,4\}}
S_{a,b} (z_{ij}) S_{a,b} (z_{jk})
S_{a,b} (z_{kl}) S_{a,b} (z_{li})\,\tr(F^{(i)}F^{(j)}F^{(k)}F^{(l)}) \label{e:Sdef}
\\
\cS_{2;a,b} &=& -{1\over2^5}
\sum_{\{(i,j),(k,l)\}=\{1,2,3,4\}}\!\!\!\!\!\!\! S_{a,b} (z_{ij})S_{a,b}
(z_{ji}) \,\tr(F^{(i)}F^{(j)}) \,
(\epsilon^{(k)}\cdot \cQ_k \, \epsilon^{(l)}\cdot \cQ_l + {1\over
2\alpha'}\epsilon^{(k)}\cdot\epsilon^{(l)}\, \cT(z_{kl}))\nn \,. 
\end{eqnarray}
Because the orbifold action only affects the right-moving
fermionic zero modes, the left movers are untouched and Riemann's identities
imply (see Appendix~\ref{sec:green-functions} for
details)
\begin{equation}
\sum_{a,b=0,1\atop ab=0} (-1)^{a+b+ab}    \cZ_{a,b}\, \cW_{a,b}=\left(\pi\over 2\right)^4\, t_8
F^4\,.
\label{e:Riemann}
\end{equation}
Notice that a contribution with less than four fermionic contractions
vanishes. 
We now rewrite \eqref{e:4grav-IIasym}:
\begin{eqnarray}
\label{eq:IIassymAmp} \cM^{(6)}_{(4,0)II}&=&-\cN\,\frac12\,\left(\pi\over2\right)^4\, t_8
F^4\,\int_{\cF}
 {d^2\tau\over\tau_2^{D- 6\over 2}}\int_{\cT}\prod_{1\leq i<j\leq4}
 {d^2\nu_i\over\tau_2}\, e^{\cQ}\, \times\\
\nn&\times&{1\over 2|G|}\sum_{    \bar a,\bar b=0,1\atop g,h}(-1)^{\bar a+\bar
    b+\mu \bar a\bar b} (-1)^{C(\bar a,\bar b,g,h)} \,\bar \cZ_{\bar
a, \bar b}^{h,g}
\bar \cW_{\bar a,\bar b} \,.
\label{e:MII40.6.partial}
\end{eqnarray}
For the class of asymmetric $\ZZ_N$ orbifolds with $n_v$ vector
multiplets studied here, the partition function $\cZ_{a,b}^{(asym)}=|G|^{-1}\,\sum_{g,h}
(-1)^{C(a,b,g,h)} \cZ_{a,b}^{g,h}$ has the following low-energy expansion:
\begin{equation}
\cZ_{0,0}^{(asym)}=\frac{1}{\sqrt q}+n_v+2+o(q)\,;\qquad \cZ_{0,1}^{(asym)}=\frac{1}{\sqrt q}-(n_v+2)+o(q)\,;\qquad \cZ_{1,0}^{(asym)}=0+o(q)\,.
\label{e:asymnv}
\end{equation}
Because the four-dimensional fermionic zero modes are not saturated we have
$\cZ^{asym}_{1,1}=0$. 

Since in those constructions no massless mode arises in the twisted
$h\neq0$ sector, this sector decouples. Hence, at $o(\bar q)$ one has the
following relation:
\begin{equation}\label{e:Asymliminter}
 \sum_{    \bar a,\bar b=0,1}(-1)^{a+b+ab}
\bar\cZ_{\bar a,\bar b}^{(asym)} \bar\cW_{\bar a,\bar b}\to 
(\bar \cW_{0,0}-\bar
\cW_{0,1})|_{\sqrt q}+(n_v+2)(\bar \cW_{0,0}+\bar
\cW_{0,1})|_{q^0}\,.
\end{equation}
The contribution of massless states to the field theory amplitude is given
by 
\begin{equation}
  \label{e:II-asymI}
  \cA^{(n_v)}_{(4,0)II}=\frac14\,\left(\pi\over2\right)^4\,t_8 F^4
  \,\left((\bar \cW_{0,0}-\bar
\cW_{0,1})|_{\sqrt q}+(n_v+2)(\bar \cW_{0,0}+\bar
\cW_{0,1})|_{q^0}\right)\,.
\end{equation}
Using the Riemann identity~\eqref{e:Riemann} we can rewrite this
expression in the following form
\begin{equation}
  \label{e:II-asym}
  \cA^{(n_v)}_{(4,0)II}=\frac14\,\left(\pi\over2\right)^4\,t_8 F^4
  \,\left(\left(\pi\over2\right)^4t_8\tilde F^4+(n_v-6)(\bar \cW_{0,0}+\bar \cW_{0,1})|_{q^0} +
16\bar \cW_{1,0}|_{q^0} \right)\,.
\end{equation}
Higher powers of $\bar q$ in $\cW_{a,b}$ or in $\cQ$
are suppressed in the field theory limit that we discuss in 
section~\ref{sec:qftlimit-one-loop}.

At this level, this expression is not identical to the one derived
in the heterotic construction~\eqref{e:amp-het}.
The type~II and
heterotic $(4,0)$ string  models with $n_v$ vector multiplets 
are dual to each other under the transformation $S\to -1/S$ where $S$
is the axion-dilaton scalar in the $\cN=4$ supergravity multiplet. 
We will see in section \ref{sec:comparing} that
for the four-graviton amplitudes we
obtain the same answer after integrating out the real parts of the positions
of the vertex operators.

We now illustrate this analysis on the examples of the asymmetric orbifold
with six or zero vector multiplets.

\subsubsection{Example: A model with six vector multiplets}
\label{sec:model-with-six}
Let us compute the partition function of the asymmetric orbifold
obtained by the action of the right-moving fermion counting operator
$(-1)^{F_R}$ and a $\ZZ_2$ action on the torus
$T^6$~\cite{Sen:1995ff,Gregori:1997hi}. The effect of the $(-1)^{F_R}$
orbifold is to project out the sixteen vector multiplets arising from the
R/R sector, while preserving supersymmetry on the right-moving
sector.  The moduli space of the theory is given by~\eqref{e:modspace} with
$n_v=6$ and $\Gamma=\Gamma(2)$ (see~\cite{Gregori:1997hi}
for instance). 

The partition function for the $(4,0)$ CFT $\ZZ_2$ asymmetric orbifold
model $\cZ_{a,b}^{(asym),(n_v=6)}=\frac 1 2  \sum_{g,h} \cZ_{a,b}^{g,h}$ with
\begin{equation}
   \cZ^{h,g}_{a,b}(w):=(-1)^{ag+bh+gh}\cZ_{a,b}\Gamma_{(4,4)}\,
\Gamma_{(2,2)}^w\left[h\atop g\right]\,, 
\label{e:ZIIasymabgh}
\end{equation}
where the shifted lattice sum $\Gamma_{(2,2)}^w\left[h\atop g\right]$ is given
in~\cite{Gregori:1997hi} and recalled in Appendix~\ref{sec:chblocks}. The
chiral blocks $\cZ_{a,b}$ have been defined in \eqref{e:chblkuntw} and
$\Gamma_{(4,4)}$ is the lattice sum of the $T^4$. Using the fact that
$\Gamma_{(2,2)}^w\left[h\atop g\right]$ reduces to $0$ for $h=1$, to $1$ for 
$h=0$ and that $\Gamma_{(4,4)}\to 1$ in the field theory limit, we see
that the partition function is unchanged in the sectors $(a,b)=(0,0)$ and
$(0,1)$ while for the $(a,b)=(1,0)$ sector, the $(-1)^{ag}$ in
\eqref{e:ZIIasymabgh} cancels the partition function when summing over $g$. One
obtains the following result :
\begin{equation}
  \cZ_{0,0}^{(asym),(n_v=6)}=\cZ_{0,0}\,;\qquad
\cZ_{0,1}^{(asym),(n_v=6)}=\cZ_{0,1}\,;\qquad
\cZ_{1,0}^{(asym),(n_v=6)}=0\,.
\label{e:ZIIasym6}
\end{equation}
Using \eqref{e:chblksexp}, one checks directly that it corresponds to
\eqref{e:asymnv} with $n_v=6$.

\subsubsection{Example: Models with zero vector multiplet}
\label{sec:models-with-zero}

Now we consider the type II asymmetric orbifold models
with zero vector multiplets constructed in~\cite{Dabholkar:1998kv} and
discussed in~\cite{Mizoguchi:2001cp}.

Those models are compactifications of the type~II superstring on a
six-dimensional torus with an appropriate choice for the value of the
metric $G_{ij}$ and B-field $B_{ij}$. The Narain lattice is given by
$\Gamma^{DH}=\{p_L,p_R;  p_L,p_R\in \Lambda_W(\mathfrak g), p_L-p_R\in
\Lambda_R(\mathfrak g)\}$ where $\Lambda_R(\mathfrak g)$ is the root
lattice of  a simply laced semi-simple Lie
algebra $\mathfrak g$, and $\Lambda_W(\mathfrak g)$ is the weight
lattice.

The asymmetric orbifold action is given by $|p_L,p_R\rangle \to
e^{2i\pi p_L \cdot v_L}\, |p_L, g_R p_R\rangle$ where $g_R$ is an element of
the Weyl group of $\mathfrak g$ and $v_L$ is a shift vector appropriately
chosen to avoid any massless states in the twisted
sector~\cite{Dabholkar:1998kv,Mizoguchi:2001cp}. With such a choice of
shift vector and because the asymmetric orbifold action leaves $p_L$
invariant, we have $(4,0)$ model of four-dimensional supergravity with no
vector
multiplets.

The partition function is given by
\begin{equation}
 \cZ^{asym}_{\bar a,\bar b}= {\theta\abb \over (\eta(\bar\tau))^3 }\,{1\over|G|}\sum_{\{g_j,h_j\}} \prod_{i=1}^3
\cZ_{\bar a,\bar b}^{h_ j,g_j}\,,
\end{equation}
where the sum runs over the sectors of the orbifold. For instance, in the
$\ZZ_9$ model of Dabholkar and Harvey, one has $g_j \in
j\times\{\frac29,\frac49,\frac89\}$ with $j=0,\dots,8$ and the same for $h_j$.
The twisted conformal blocks are:

\begin{equation}
 \cZ_{\bar a,\bar b}^{h,g}=
\left\{
  \begin{array}{llr}
    \left({\theta \abb \over \eta(\bar \tau)}\right) ^3 \times \left({1\over
\eta(\bar\tau)}\right)^6 &\mathrm{if}\ (g,h)=(0,0) \mod 2,\\
    e^{i {\pi\over2} a(g-b)}2 \sin(\frac {\pi g} 2) {\theta \left[ a+h \atop b+g
\right] \over \theta \left[ 1+h \atop 1+g \right]} &\forall (g,h)\neq(0,0) \mod 2.
  \end{array}
\right.
\end{equation}
The phase in~\eqref{e:4grav-IIasym} is determined by modular
invariance to be
$ C(\bar a,\bar b,  g_R,h_R)=\sum_{i=1}^3 (\bar a g^i_R +\bar b h^i_R+
  g^i_R h^i_R)$.

In the field theory limit, we perform the low-energy expansion of this
partition function and we find that it takes the following form for all of the
models in~\cite{Dabholkar:1998kv,Mizoguchi:2001cp}:
\begin{equation}
\cZ_{0,0}^{(asym),(n_v=0)}=\frac{1}{\sqrt q}+2+o(q)\,;\quad
\cZ_{0,1}^{(asym),(n_v=0)}=\frac{1}{\sqrt q}-2+o(q)\,;\quad
\cZ_{1,0}^{(asym),(n_v=0)}=0+o(q)\,,
\label{e:asymnv0}
\end{equation}
which is \eqref{e:asymnv} with $n_v=0$ as expected.

\subsection{Type~II symmetric orbifold}
\label{sec:one-loop-sym}

In this section we consider $(2,2)$ models of four-dimensional $\cN=4$ supergravity. These models can be obtained from the compactification of
type~II string theory on symmetric orbifolds of $K_3\times
T^2$. The difference with
the heterotic models considered in section~\ref{sec:one-loop-CHL} is that the
scalar parametrizing the coset space $SU(1,1)/U(1)$ that used to be the
axio-dilaton $S$ is now the K\"ahler modulus of
the two-torus $T^2$ for the type~IIA case or complex structure modulus for the
type~IIB case. The non-perturbative duality
relation between these two models is discussed in detail
in~\cite{Gregori:1997hi,Aspinwall:1995fw}.

Models with $n_v\geq2$ have been constructed
in~\cite{Ferrara:1989nm}.  
The model with $n_v=22$ is a  $T^4/\ZZ_2\times T^2$ orbifold, and the following
models with $n_v\in\{14,10\}$ are successive $\ZZ_2$ orbifolds of the first
one. The model with $n_v=6$ is a freely acting $\ZZ_2$ orbifold of the
$T^4/\ZZ_2\times T^2$ theory that simply projects out the sixteen vector
multiplets of the R/R sector.
The four-graviton amplitude can be effectively written in terms of the
$(g,h)$ sectors of the first $\ZZ_2$ orbifold of the $T^4$, and
writes
\begin{eqnarray}
&& \label{e:4grav-II-22eff} \cM^{(n_v)}_{(2,2)}=\cN\,\int_{\cF}
 {d^2\tau\over\tau_2^{D- 6\over 2}}\int_{\cT}\prod_{1\leq i<j\leq4}
 {d^2\nu_i\over\tau_2}\, e^{\mathcal Q}\, \times\\
\nn&\times&{1\over 4|G|}\sum_{h,g=0}^1\, \sum_{a,b=0,1\atop  \bar a,\bar
b=0,1} (-1)^{a+b+ab} (-1)^{\bar a +\bar b +\bar a \bar b}
\cZ_{a, b}^{h,g,(n_v)} \bar \cZ_{\bar a,\bar b}^{h,g,(n_v)}\,
 (\cW_{ a,b}\bar \cW_{\bar a,\bar b}+\cW_{a,b;\bar a, \bar b})\,,
\end{eqnarray}
where $\cN$ is the same overall normalization as for the previous
amplitudes and $\cZ^{h,g,(n_v)}_{a,b}$ is defined in Appendix~\ref{sec:chblocks}.
The term $\cW_{a,b;\bar a,\bar b}$ is a mixed term made of
contractions between holomorphic and anti-holomorphic fields. It does not
appear in the $(4,0)$ constructions since the left/right contractions
vanish due to the totally unbroken supersymmetry in the left-moving sector.

Two types of contributions arise from the mixed correlators

\begin{eqnarray}
\cW^1_{a,b;\bar a,\bar b}&=&  {\langle :\epsilon^{(i)}\cdot\partial X
    \tilde\epsilon^{(i)}\cdot\bar\partial X: :\epsilon^{(j)}\cdot\partial X
    \tilde\epsilon^{(j)}\cdot\bar\partial X:  \,\prod_{r=1}^4 e^{i k_r\cdot X(z_r)}
    \rangle_{a,b;\bar a,\bar b}\over
(2\alpha')^4\langle \prod_{j=1}^4 e^{i k_j\cdot X(z_j)}
\rangle};\\
\nn\cW^2_{a,b;\bar a,\bar b}&=&{\langle \epsilon^{(i)}\cdot\partial X\,\epsilon^{(j)}\cdot\partial X
    \tilde\epsilon^{(k)}\cdot\bar\partial X\,
    \tilde\epsilon^{(l)}\cdot\bar\partial X \,\prod_{r=1}^4 e^{i k_r\cdot X(z_r)}
    \rangle_{a,b;\bar a,\bar b}\over
(2\alpha')^4\langle \prod_{j=1}^4 e^{i k_j\cdot X(z_j)}
\rangle}\,,\qquad (ij)\neq (kl)
\end{eqnarray}
with at least one operator product expansion (OPE) between a holomorphic and an anti-holomorphic operator.
Explicitely, we find
\begin{eqnarray}
\label{e:WLR}
\cW_{a,b;\bar a;\bar
  b}&=&\sum_{\{i,j,k,l\}\in\{1,2,3,4\}\atop (i,j)\neq (k,l)}
(S_{a,b}(\nu_{ij}))^2 (\bar S_{\bar a,\bar
  b}(\bar\nu_{kl}))^2\,\times\tr(F^{(i)}F^{(j)}) \tr(F^{(k)}F^{(l)})\cr
&\times&\,\left( \epsilon^{(k)}\cdot\tilde\epsilon^{(i)} \hat \cT(k,i)
   (\epsilon^{(l)} \cdot \tilde \epsilon^{(j)} \hat \cT(l,j)+\epsilon^{(l)}
   \cdot\cQ_l \ \tilde \epsilon^{(j)} \cdot \bar\cQ_j) +(i\leftrightarrow
   j)\right.\cr
&+&\left.\epsilon^{(k)} \cdot\cQ_k (\epsilon^{(l)}
\cdot\tilde\epsilon^{(i)}\ \tilde\epsilon^{(j)} \cdot \bar\cQ_j 
\hat\cT(l,i)+ \epsilon^{(l)} \cdot\tilde\epsilon^{(j)} \hat\cT(l,j)
\tilde\epsilon^{(i)} \cdot\bar\cQ_i )\right)\cr
&+&\sum_{\{i,j,k,l\}\in\{1,2,3,4\}} |S_{a,b}(\nu_{ij})|^4 \times(\tr(F^{(i)}F^{(j)}))^2\cr
&\times&\,\left( \epsilon^{(k)}\cdot\tilde\epsilon^{(l)} \hat \cT(k,l)
  (\epsilon^{(l)} \cdot \tilde \epsilon^{(k)} \hat \cT(l,k)+\epsilon^{(l)}
   \cdot\cQ_l \ \tilde \epsilon^{(k)} \cdot \bar\cQ_k) +(k\leftrightarrow
   l)\right)
\end{eqnarray}
where
\begin{equation}
  \label{e:cThat}
\hat \cT(i,j):=\partial_{\nu_i}\bar \partial_{\bar\nu_j}  \cP(\nu_i-\nu_j|\tau)={\pi\over4}\left({1\over\tau_2}-\delta^{(2)}(\nu_i-\nu_j)\right)\,.
\end{equation}
Forgetting about the lattice sum, which at any rate is equal to one in
the field theory limit, 
\begin{equation}
\cZ^{h,g}_{a,b}=c_h
{(\theta\left[a\atop
b\right](0|\tau))^2 \theta\left[a+h\atop b+g\right](0|\tau)\theta\left[a-h\atop
b-g\right](0|\tau)\over (\eta(\tau))^6(\theta\left[ 1+h \atop 1+g \right](0|\tau))^2}
=c_h(-1)^{(a+h)g}\, \left({\theta\left[a\atop
b\right](0|\tau) \theta\left[a+h\atop b+g\right](0|\tau)\over (\eta(\tau))^3\theta\left[ 1+h \atop 1+g \right](0\tau)}\right)^2\,,
\label{e:ZIIsym}
\end{equation}
where $c_h$ is an effective number whose value depends on $h$ in the following
way: $c_{0}=1$ and $c_{1}=\sqrt{n_v-6}$. This number represents the successive halving of the number of twisted R/R states.
We refer to Appendix~\ref{sec:chblocks} for details.

The sum over the spin structures in the untwisted sector $(g,h)=(0,0)$ is
once again performed using Riemann's identities:
\begin{equation}
 \sum_{a,b=0,1\atop ab=0} (-1)^{a+b+ab} \cZ_{a,b}^{0,0}
\cW_{a,b}=\left(\pi\over2\right)^4\, t_8F^4\,. 
\end{equation}
In the twisted sectors $(h,g)\neq (0,0)$ we remark that
 $\cZ^{0,1}_{0,1}=Z^{0,1}_{0,0}$,
$\cZ^{1,0}_{1,0}=\cZ^{1,0}_{0,0}$,  $\cZ^{1,1}_{1,0}=\cZ^{1,1}_{0,1}$, and 
$   \cZ^{1,0}_{0,1}=\cZ^{0,1}_{1,0}=\cZ^{1,1}_{0,0}=0$, which gives for the
chiral blocks in~\eqref{e:4grav-II-22eff}: \begin{eqnarray}
 \sum_{a,b=0,1\atop ab=0} (-1)^{a+b+ab} \cZ_{a,b}^{0,1}
\nn \cW_{a,b}&=&\cZ_{0,0}^{0,1}\,\left(\cW_{0,0}-\cW_{0,1}\right),\\
 \sum_{a,b=0,1\atop ab=0} (-1)^{a+b+ab} \cZ_{a,b}^{1,0}
 \cW_{a,b}&=&\cZ_{0,0}^{1,0}\,\left(\cW_{0,0}-\cW_{1,0}\right),\\
 \nn\sum_{a,b=0,1\atop ab=0} (-1)^{a+b+ab} \cZ_{a,b}^{1,1}
 \cW_{a,b}&=&\cZ_{0,1}^{1,1}\,\left(\cW_{0,1}-\cW_{1,0}\right)\,.
\end{eqnarray}
\\Therefore the factorized terms in the correlator take the simplified
form
\begin{multline}
{1\over 4|G|}\sum_{   g,h}\, \sum_{a,b=0,1\atop  \bar a,\bar b=0,1}   (-1)^{a+b+ab}
(-1)^{\bar a +\bar
b +\bar a \bar b} 
\cZ_{a, b}^{h,g} \bar \cZ_{\bar a,\bar b}^{ h, g}\,
 \cW_{ a,b}\bar \cW_{\bar a,\bar b}\cr
=\frac18\,
\left(\pi\over 2\right)^8 t_8t_8R^4 + \frac18\, \left|\cZ^{0,1}_{0,0}
(\cW_{0,0}-\cW_{0,1})\right|^2
+\frac18\,\left|\cZ^{1,0}_{0,0}
(\cW_{0,0}-\cW_{1,0})\right|^2+\frac18\,\left|\cZ^{1,1}_{0,1}
(\cW_{0,1}-\cW_{1,0})\right|^2 \,,
\label{e:amp221l}
\end{multline}
where $ t_8t_8R^4$ is the Lorentz scalar built from four powers of
the Riemann tensor arising at the linearized level as the product
$ t_8t_8R^4 =t_8F^4\,t_8\tilde F^4$.
\footnote{ This Lorentz  scalar  is the one
obtained from the four-graviton tree amplitude $t_8t_8R^4=stu
M^{tree}(1,2,3,4)$ setting Newton's constant to one.}

The mixed terms can be treated in the same way with the result
\begin{multline}
    {1\over 4|G|}\sum_{   g,h}\, \sum_{a,b=0,1\atop  \bar a,\bar b=0,1}   (-1)^{a+b+ab}
(-1)^{\bar a +\bar
b +\bar a \bar b} 
\cZ_{a, b}^{h,g} \bar \cZ_{\bar a,\bar b}^{ h, g}\,
 \cW_{ a,b;\bar a,\bar b}\cr
= \frac18\, |\cZ^{0,1}_{0,0}|^2
(\cW_{0,0;0,0}-\cW_{0,1;0;1})
+\frac18\,|\cZ^{1,0}_{0,0}|^2
(\cW_{0,0;0,0}-\cW_{1,0;1,0})+\frac18\,|\cZ^{1,1}_{0,1}|^2
(\cW_{0,1;0,1}-\cW_{1,0;1,0})\,,
\label{e:amp22LR}
\end{multline}

Since the conformal blocks $\cZ^{h,g}_{a,b}$ have the $q$ expansion (see
\eqref{e:thetaq}) %
\begin{equation}\label{e:Zq}
  \cZ^{0,1}_{0,0}={1\over\sqrt q}+4\sqrt q+o(q)  \,;~
  \cZ^{1,0}_{0,0}= 4\sqrt{n_v-6}+o(q) \,;~
  \cZ^{1,1}_{0,1}=4\sqrt{n_v-6}+o(q)\,,
\end{equation}
the massless  contribution to the integrand of \eqref{e:amp221l} is given by 
\begin{multline}\label{e:AII22}
\cA_{(2,2)}^{(n_v)}=\frac18\,\left(\pi\over2\right)^8 t_8t_8R^4 +\frac18\,
\Big|\cW_{0,0}|_{\sqrt q}-\cW_{0,1}|_{\sqrt q}\Big|^2+\frac18\, (\cW_{0,0;0,0}|_{\sqrt q}-\cW_{0,1;0,1}|_{\sqrt q})\cr
+ 2(n_v-6)\,
\Big(\Big|\cW_{0,0}|_{q^0}-\cW_{1,0}|_{q^0}\Big|^2+\Big|\cW_{0,1}|_{q^0}
-\cW_{1,0}|_{q^0}\Big|^2\Big)\cr
+ 2(n_v-6)\,\Big(\cW_{0,0;0,0}|_{q^0;\bar q^0}+\cW_{0,1;0,1}|_{q^0;\bar
q^0}-2\cW_{1,0;1,0}|_{q^0;\bar q^0}\Big) \,.
\end{multline}
We notice that the bosonic piece $\cW^B$  in
$\cW_{a,b}$ in~\eqref{e:Wabdef} cancels in each term of the previous
expression, due to the minus sign between the $\cW_{a,b}$'s in the squares.

The integrand of the four-graviton amplitude takes a different form in
the $(2,2)$ construction compared with the expression for the $(4,0)$
constructions in heterotic in~\eqref{e:amp-het} and asymmetric type~II
models in~\eqref{e:II-asym}. We will show  that after taking the field theory limit and
performing the integrals the amplitudes will turn out to be the same.

As mentioned above, for a given number of vector multiplets the type II $(2,2)$
models are only non-perturbatively equivalent (U-duality) to the $(4,0)$
models. However, we will see that this non-perturbative duality does not affect
the perturbative one-loop multi-graviton amplitudes. Nevertheless, we expect
that both $\alpha'$ corrections to those amplitudes and amplitudes with external
scalars and vectors should be model dependent. 

In the next section, we analyze
the relationships between the string theory models.

\section{Comparison of the string models}
\label{sec:comparing}
\subsection{Massless spectrum}
\label{sec:massl-spectr}

The spectrum of the type~II superstring in ten dimensions is given by the following
GSO sectors: the graviton $G_{MN}$, the $B$-field
$B_{MN}$, and the dilaton $\Phi$ come from the NS/NS sector, the gravitini
$\psi^M$, and the dilatini $\lambda$ come from the R/NS and NS/R sectors,
while the one-form $C_M$ and three-form  $C_{MNP}$ come from the R/R sector
in the type  IIA string.
The dimensional reduction of type II string on a torus preserves all of the
thirty-two supercharges and leads to the $\cN=8$ supergravity multiplet in four
dimensions.

The reduction to $\cN=4$ supersymmetry preserves sixteen supercharges and leads
to the following content.
The  NS/NS sector contributes to the $\cN=4$ supergravity multiplet
 and to six vector multiplets.
The R/R sector contributes to the $\cN=4$ spin-$\frac32$
  multiplet and to the vector multiplets with a multiplicity depending on the
model.

In the partition function, the first Riemann vanishing identity 
\begin{equation}\label{e:R1}
  \sum_{a,b=0,1} (-1)^{a+b+ab} \cZ_{a,b} =0  \,,
\end{equation}
reflects the action of the $\cN=4$ supersymmetry inside the one-loop amplitudes
in the following manner. The $q$ expansion of this identity gives 
\begin{equation}
  \left({1\over \sqrt q}+ 8+o(\sqrt q)\right)-     \left({1\over \sqrt q}- 8+o(\sqrt
q)\right)-     \left(16+o(q)\right)=0\,.
\end{equation}
The first two terms  are the expansion of $\cZ_{0,0}$ and $\cZ_{0,1}$ and
the last one is the expansion of $\cZ_{1,0}$.
 The cancellation of the $1/\sqrt q$ terms shows that the GSO projection
eliminates the tachyon from the spectrum, and at the order $q^0$ the
cancellation results in the matching between the bosonic and
fermionic degrees of freedom.

In the amplitudes, chiral $\cN=4$ supersymmetry implies the
famous Riemann identities, stating that for $0\leq n\leq 3$ external legs,
the one-loop $n$-point amplitude vanishes  (see eq.~\eqref{e:Riemann03}). At
four points it gives:
\begin{equation}\label{e:gsoN4}
  \sum_{a,b=0,1} (-1)^{a+b+ab} \cZ_{a,b} \cW_{a,b}= \left(\pi\over2\right)^4 t_8F^4  \,.
\end{equation}
In $\cW_{a,b}$, see~\eqref{e:WFabdef}, the term independent of the spin
structure $\cW^B$ and the terms with less than four
fermionic contractions $\cS_{2;a,b}$ cancel in the previous identity. The
cancellation of the tachyon yields at the first order in the $q$ expansion
of~\eqref{e:gsoN4} \begin{equation}
  \cW_{0,0}|_{q^0} -  \cW_{0,1}|_{q^{0}}=0\,.
\label{e:gsonotach}
\end{equation}
The next term in the expansion gives an identity describing the propagation
of the $\cN=4$ super-Yang-Mills multiplet in the loop
\begin{equation}
8(\cW_{0,0}|_{q^0} +   \cW_{0,1}|_{q^0}-2
\cW_{1,0}|_{q^0})+(\cW_{0,0}|_{\sqrt q}
- \cW_{0,1}|_{\sqrt q}) =\left(\pi\over2\right)^4 t_8F^4\,.
\label{e:gsoym}
\end{equation}
In this equation, one should have vectors, spinors and scalars
propagating according to the sector of the theory. In $\cW_{a,b}$,
$a=0$ is the NS  sector, and $a=1$ is the Ramond sector.
 The
scalars have already been identified in \eqref{e:gsonotach} and correspond to
$\cW_{0,0}|_{q^0}+\cW_{0,1}|_{q^0}$. The vector, being a
massless bosonic
degree of freedom should then correspond to $\cW_{0,0}|_{\sqrt
q}-\cW_{0,1}|_{\sqrt q}$. Finally, the fermions correspond to
$\cW_{1,0}|_{q^0}$.
The factor of eight in front of the first term is the number of degrees of freedom of a
vector in ten dimensions; one can check that the number of bosonic degrees of
freedom matches the number of fermionic degrees of freedom.

\subsection{Amplitudes and supersymmetry}
\label{sec:ampl-contr}

In this section we discuss the relationships between the four-graviton amplitudes in
the various $\cN=4$ supergravity models in the field theory limit. We apply the
logic of the previous section about the spectrum of left or right movers to
the tensor product spectrum and see that we can precisely identify the
contributions to the amplitude, both in the $(4,0)$ and $(2,2)$ models. The
complete evaluation of the amplitudes will be performed in
section~\ref{sec:qftlimit-one-loop}.

As mentioned above, the field theory limit is obtained by considering the large
$\tau_2$ region, and the integrand of the field theory amplitude is given by 

\begin{equation}
 A^{(n_v)}_X= \int_{-\frac12
 }^{\frac12}\prod_{i=1}^4 d\nu_i^1 \, \cA^{(n_v)}_X \,,
\end{equation}
where  $X\in\{(4,0)het,\,(4,0)II,\,(2,2) \}$ indicates the model, as
in~\eqref{e:amp-het},~\eqref{e:II-asym} or~\eqref{e:AII22} respectively.

At one loop this quantity  is the sum of the contribution from  $n_v$  $\cN=4$
vector (spin-1) supermultiplets running in the loop and the $\cN=4$ spin-2 supermultiplet 
\begin{equation}
     A^{(n_v)}_X= A^{spin-2}_X+ n_v\, A^{spin-1}_X\,.
\end{equation}

For the case of the type~II asymmetric orbifold models with $n_v$ vector
multiplets we deduce from~\eqref{e:II-asym}
\begin{equation}\label{e:IIvm}
A^{spin-1}_{(4,0)II} = \frac14\,\left(\pi\over2\right)^4\, t_8F^4 \,   \int_{-\frac12
 }^{\frac12}\prod_{i=1}^4 d\nu_i^1 \,    (\cW_{0,0}|_{q^0}+ \cW_{0,1}|_{q^0})\,.
\end{equation}
Since $t_8F^4$ is the supersymmetric left-moving sector
contribution (recall the supersymmetry identity in~\eqref{e:gsoN4}), it
corresponds to an
$\cN=4$ vector multiplet and
 we recognize in \eqref{e:IIvm} the product of this multiplet with the scalar 
from the right-moving sector:
\begin{equation}
    (1_{\bf 1}, 1/2_{\bf 4},0_{\bf 6})_{\cN=4}=
 (1_{\bf 1}, 1/2_{\bf 4},0_{\bf 6})_{\cN=4}\otimes (0_{\bf 1})_{\cN=0}\,.
\end{equation}
This agrees with the identification made
in the previous subsection where $\cW_{0,0}|_{q^0}+\cW_{0,1}|_{q^0}$ was argued
to be a scalar contribution.

The contribution from the  $\cN=4$ supergravity multiplet running in the loop is
given by
\begin{equation}
A^{spin-2}_{(4,0)II} = \frac14\,\left(\pi\over2\right)^4\, t_8F^4  \, \int_{-\frac12
 }^{\frac12}\prod_{i=1}^4 d\nu_i^1 \,  \Big[ 2  (\cW_{0,0}|_{q^0}+
 \cW_{0,1}|_{q^0})+  (\cW_{0,0}|_{\sqrt q}- \cW_{0,1}|_{\sqrt q})\Big]\,.
\label{e:IIspin2}
\end{equation}
The factor of 2 is the number of degrees of freedom of a vector in
four dimensions. Since $\cZ^{asym}_{1,0}=0+o(q)$ for the $(4,0)$ model
asymmetric orbifold construction, the integrand of the four-graviton amplitude
in~\eqref{e:asymnv} does not receive any contribution from the right-moving R
sector. Stated differently, the absence of $\cW_{1,0}$ implies that both R/R and
NS/R sectors are projected out, leaving only the contribution
from the NS/NS and R/NS. Thus, the four $\cN=4$ spin-$\frac32$ supermultiplets
and sixteen $\cN=4$ spin-1 supermultiplets are projected out, leaving
at most six  vector multiplets. This number is further reduced to zero  in the
Dabholkar-Harvey construction~\cite{Dabholkar:1998kv}.

From~\eqref{e:IIspin2} we recognize that the $\cN=4$ supergravity multiplet is
obtained by the following tensor product 
\begin{equation}
     (2_{\bf 1}, 3/2_{\bf 4},1_{\bf 6} , 1/2_{\bf 4},0_{\bf
       2})_{\cN=4}=
 (1_{\bf 1}, 1/2_{\bf 4},0_{\bf
   6})_{\cN=4}\otimes 
 (1_{\bf 1})_{\cN=0}\,.
\end{equation}
The two real scalars arise from the trace part and the anti-symmetric part
(after dualisation in four dimensions) of the tensorial product of the
two vectors. Using the identification of
$\cW_{0,0}|_{q^0}+\cW_{0,1}|_{q^0}$ with a scalar
contribution and the equation~\eqref{e:Wabrel} we can now identify 
$ \cW_{0,0}|_{\sqrt q}-\cW_{0,0}|_{\sqrt q}$
with the contribution of a vector and two scalars.
This confirms the identification of $\cW_{1,0}|_{q^0}$ with a spin-$\frac12$
contribution in the end of section~\ref{sec:massl-spectr}.

Since 
\begin{equation}
  (3/2_{\bf 1},1_{\bf 4}, 1/2_{\bf 6+1},0_{\bf 4+\bar 4})_{\cN=4}
= (1_{\bf 1}, 1/2_{\bf 4},0_{\bf 6})_{\cN=4}\otimes 
 (1/2)_{\cN=0}\,,
\end{equation}
we see that removing the four spin $\frac12$ (that is, the term
$\cW_{1,0}|_{q^0}$)
 of the right-moving massless spectrum of the string theory
construction in asymmetric type~II models removes the contribution from the
massless spin $\frac32$ to the amplitudes. For the asymmetric type~II model,
using~\eqref{e:gsoym},  we can present the contribution from the $\cN=4$ 
supergravity multiplet in a form that reflects the
decomposition of the $\cN=8$ supergravity multiplet into $\cN=4$ supermultiplets
\begin{eqnarray}
  (2_{\bf 1}, 3/2_{\bf 8},1_{\bf 28} , 1/2_{\bf 56},0_{\bf 70})_{\cN=8}&=&   
 (2_{\bf 1}, 3/2_{\bf 4},1_{\bf 6} ,1/2_{\bf 4},0_{\bf 1+1})_{\cN=4}\nn\\
\label{e:multdec} &\oplus&\,{\bf 4}\,
 (3/2_{\bf 1},1_{\bf 4}, 1/2_{\bf 6+1},0_{\bf 4+\bar 4})_{\cN=4}\\
\nn &\oplus& {\bf 6}\,
 (1_{\bf 1},1/2_{\bf 4},0_{\bf 6})_{\cN=4}\,, 
\end{eqnarray}
as
\begin{equation}\label{e:40spin2}
A_{(4,0)II}^{spin-2}= A^{spin-2}_{\cN=8} -
6\,A^{spin-1}_{(4,0)II}-4\, A^{spin-\frac32}_{(4,0)II}\,,
\end{equation}
where we have introduced the $\cN=8$  spin-2 supergravity contribution 
\begin{equation}
  \label{e:Amp2N8}
A^{spin-2}_{\cN=8}=\frac14 \, \left(\pi\over2\right)^8\, t_8t_8R^4  \,,
\end{equation}
and the  $\cN=4$  spin-$\frac32$ supergravity contribution 
\begin{equation}
  \label{e:Amp32}
A^{spin-\frac32}_{(4,0)II}=- \left(\pi\over2\right)^4\, t_8F^4  \,
\int_{-\frac12
 }^{\frac12}\prod_{i=1}^4 d\nu_i^1 \, \cW_{1,0}|_{q^0}\,.
\end{equation}

For the $(2,2)$ models the contribution of the massless states to the
amplitude  is given in~\eqref{e:AII22}. The contribution from a vector
multiplet is
\begin{multline}
  \label{e:22vm}
   A^{spin-1}_{(2,2)}= 2 \int_{-\frac12 }^{\frac12}\prod_{i=1}^4
d\nu_i^1 \, \left(\Big|\cW_{0,0}|_{q^0}- \cW_{1,0}|_{q^0}\Big|^2+
\Big|\cW_{0,1}|_{q^0}- \cW_{1,0}|_{q^0}\Big|^2\right)\cr
+ 2 \int_{-\frac12 }^{\frac12}\prod_{i=1}^4
d\nu_i^1 \, (\cW_{0,0;0,0}|_{q^0;\bar q^0}+\cW_{0,1;0,1}|_{q^0;\bar q^0}-2
\cW_{1,0;1,0}|_{q^0;\bar q^0})\,. 
\end{multline}
Using  that $\big|\cW_{0,0}|_{q^0}-
\cW_{1,0}|_{q^0}\big|^2=\big|\cW_{0,1}|_{q^0}-
\cW_{1,0}|_{q^0}\big|^2=\frac14\big|\cW_{0,0}|_{q^0}+\cW_{0,1}|_{q^0}-2
\cW_{1,0}|_{q^0}\big|^2$ as a consequence of~\eqref{e:gsonotach},
we can rewrite this as
\begin{multline}
  \label{e:vm22bis}
  A^{spin-1}_{(2,2)}= 
  \int_{-\frac12}^{\frac12} \prod_{1\leq i<j\leq 4} d\nu^1_i \,\big|\cW_{0,0}|_{q^0}+\cW_{0,1}|_{q^0}-2
\cW_{1,0}|_{q^0}\big|^2\cr
+2 \int_{-\frac12 }^{\frac12}\prod_{i=1}^4
d\nu_i^1 \, (\cW_{0,0;0,0}|_{q^0;\bar q^0}+\cW_{0,1;0,1}|_{q^0;\bar q^0}-2
\cW_{1,0;1,0}|_{q^0;\bar q^0})\,,
\end{multline}
 showing that this spin-1 contribution in the $(2,2)$ models arises as
  the product of two $\cN=2$  hypermultiplets
$Q=(2\times 1/2_{\bf 1},2\times 0_{\bf 2})_{\cN=2}$  
\begin{equation}
2\times(1_{\bf 1},1/2_{\bf 4} ,0_{\bf 6})_{\cN=4}= (2\times 1/2_{\bf
  1},2\times 0_{\bf 2})_{\cN=2}\otimes (2\times 1/2_{\bf 1},2\times
0_{\bf \bar 2})_{\cN=2}\,.
\end{equation}
The contribution from  the $\cN=4$ supergravity multiplet running in
the loop (obtained  from~\eqref{e:AII22} by setting $n_v=0$) can be presented in a form reflecting the
decomposition in~\eqref{e:multdec}
\begin{equation}\label{e:22spin2}
  A^{spin-2}_{(2,2)}=A^{spin-2}_{\cN=8}-6 
A^{spin-1}_{(2,2)}-4\, A^{spin-\frac32}_{(2,2)}\,,
\end{equation}
where $A^{spin-\frac32}_{(2,2)}$ is given by
\begin{equation}
  \label{e:22spin32}
  A^{spin-\frac32}_{(2,2)}=-\frac18\,A^{spin-2}_{\cN=8}- {1\over32}\, 
  \int_{-\frac12}^{\frac12} \prod_{1\leq i<j\leq 4} d\nu^1_i
  \,\big|\cW_{0,0}|_{\sqrt q}-\cW_{0,1}|_{\sqrt q}\big|^2\,.
\end{equation}
\subsection{Comparing of the string models}
\label{sec:comp-string-models}

The integrands of the amplitudes in the two $(4,0)$ models in~\eqref{e:amp-het}
and~\eqref{e:II-asym} and the $(2,2)$ models in~\eqref{e:AII22}  take a
different form. In this section we show first the equality between the
integrands  of the $(4,0)$ models and then that any difference with the $(2,2)$
models can be attributed to the contribution of the vector multiplets.

The comparison is done in the field theory limit where
$\tau_2\to+\infty$  and $\alpha'\to0$ with $t=\alpha'\tau_2$ held
fixed. The real parts of the $\nu_i$ variables are
integrated over the range $-\frac12\leq\nu_i^1\leq\frac12$.
In this limit the position of the vertex operators scale as $\nu_i= \nu_i^1
+i\tau_2 \omega_i$. The positions of the external legs  on the loop are then
denoted by $0\leq\omega_i\leq1$ and are ordered according to the kinematical
region under consideration. In this section we discuss the integration
over the $\nu_i^1$'s only; the integration over the $\omega_i$'s will be
performed in
section~\ref{sec:qftlimit-one-loop}.

\subsubsection{Comparing the $(4,0)$ models}
\label{sec:4-0-models}

In the heterotic string amplitude~\eqref{e:amp-het}, we can identify two
distinct contributions; $n_v$  vector multiplets and one $\cN=4$ supergravity
multiplet running in the loop.
At the leading order in $\alpha'$, the contribution of the vector multiplets is
given by:
\begin{equation}
\label{e:Ahetspin1}
A^{spin-1}_{(4,0)het}=\frac12\,\left(\pi\over2\right)^4\,t_8F^4\, 
\int_{-\frac12 }^{\frac12}\prod_{i=1}^4 d\nu_i^1 \,\bar\cW^B|_{\bar q^0}\,,
\end{equation}
and the one of the supergravity multiplet by
\begin{equation}
\label{e:Ahetspin2} 
 A^{spin-2}_{(4,0)het}=\frac12\,\left(\pi\over2\right)^4\,t_8F^4 \, \int_{-\frac12
 }^{\frac12}\prod_{i=1}^4 d\nu_i^1 \, \Big(
(\bar\cW^B|_{\bar q} (1+\alpha'\delta Q) + \bar\cW^B|_{\bar q^0}\cQ|_{\bar q}+ 2 \bar\cW^B|_{\bar q^0}\Big)\,.
\end{equation}
The vector multiplet contributions take  different forms in the
heterotic construction in~\eqref{e:Ahetspin1} and the type~II models in~\eqref{e:IIvm}. 
However using the expansion of the fermionic propagators given in
Appendix~\ref{sec:green-functions}, it is not difficult to perform the
integration over $\nu_i^1$  in~\eqref{e:IIvm}. We see that
\begin{equation}
\int_{-\frac12}^{\frac12}\prod_{1\leq i<j\leq4} d\nu^1_i\,  (\cW^F_{0,0}|_{\bar 
  q^0} +\cW^F_{0,1}|_{\bar q^0})=0\,.
\end{equation}
Thus there only remains the bosonic part of $\cW_{a,b}$, and we find that the contribution of the vector
multiplet is the same in
the heterotic and asymmetric orbifold constructions 
\begin{equation}\label{e:40spin1eq}
  A^{spin-1}_{(4,0)het}= A^{spin-1}_{(4,0)II}\,.  
\end{equation}

The case of the $\cN=4$ super-graviton is a little more involved. In order to
simplify the argument we make the following choice of helicity to deal with 
more manageable expressions: $(1^{++},2^{++},3^{--},4^{--})$. We set as well
the reference
momenta $q_i$'s for graviton $i=1,\cdots,4$ as follows: $q_1=q_2=k_3$ and
$q_3=q_4=k_1$. At four points in supersymmetric theories, amplitudes with more
$+$ or $-$ helicity states vanish. In that manner the covariant quantities $t_8
F^4$
and $t_8 t_8 R^4$ are written in the spinor helicity
formalism\footnote{
A null vector $k^2=0$ is parametrized as $k_{\alpha\dot\alpha}=
k_{\alpha} \bar k_{\dot\alpha}$ where $\alpha,\dot\alpha=1,2$ are
$SL(2,\IC)$ two-dimensional spinor indices.
The positive and negative helicity polarization 
vectors are given by 
$  \epsilon^+(k,q)_{\alpha\dot\alpha}:= {q_\alpha \bar
    k_{\dot\alpha}\over\sqrt2\, \an[q,k]}$ and $\epsilon^-(k,q)_{\alpha\dot\alpha}:=- {k_\alpha \bar
    q_{\dot\alpha}\over\sqrt2\, \sq[q,k]}$, respectively,
where $q$ is a massless reference momentum.
The self-dual and anti-self-dual field strengths read 
$
  F^-_{\alpha\beta}:=\sigma^{mn}_{\alpha\beta} F_{mn}= {k_\alpha k_\beta\over\sqrt2}$ and $F^+_{\dot\alpha\dot\beta}:=\bar\sigma^{mn}_{\dot\alpha\dot\beta}F_{mn}=-
  {\bar k_{\dot\alpha} \bar k_{\dot\beta}\over\sqrt2}$, respectively.
} $2t_8F^4=\an[k_1,k_2]^2\sq[k_3,k_4]^2$,
and $4t_8t_8R^4=\an[k_1,k_2]^4\sq[k_3,k_4]^4$, respectively.  With this choice of
gauge $\epsilon^{(1)}\cdot \epsilon^{(k)}=0$ for $k=2,3,4$, $\epsilon^{(3)}\cdot
\epsilon^{(l)}=0$ with $l=2,4$ and only
$\epsilon^{(2)}\cdot\epsilon^{(4)}\neq0$. The same relationships hold for the
scalar product between the
right-moving $\tilde\epsilon$ polarizations and the left- and right-moving polarizations .
We can now simplify the various kinematical factors $\cW^B$ for the heterotic
string and the $\cW_{a,b}$'s for the type II models. We find $\bar\cW^B=\frac 12 t_8\tilde F^4\, \tilde\cW^B$ where
\begin{equation}
\tilde\cW^B= 
\tilde\cW^B_1+ {1\over \alpha' u} \tilde\cW^B_2\,,
\label{e:Wg}
\end{equation}
with
\begin{eqnarray}
\nn 
 \tilde\cW^B_1&=& \, 
    (\bar\partial \cP(12)-\bar\partial \cP(14))
    (\bar\partial \cP(21)-\bar\partial \cP(24))
    (\bar\partial \cP(32)-\bar\partial \cP(34))
    (\bar\partial \cP(42)-\bar\partial \cP(43))\,,\\
 \tilde\cW^B_2&=& \bar\partial^2 \cP(24) 
 (\bar\partial \cP(12)-\bar\partial \cP(14))
 (\bar\partial \cP(32)-\bar\partial \cP(34))\,.
 \label{e:Wg_i}
\end{eqnarray}
In these equations it is understood that $\cP(ij)$ stands for
$\cP(\nu_i-\nu_j)$.
We find as well that $\cW_{a,b}^F=\frac 12 t_8 F^4\, \tilde\cW^F_{a,b}$ with
$\tilde
\cW_{a,b}^F =\tilde  \cS_{4;a,b}+ \tilde \cS_{2;a,b}$,
where
\begin{eqnarray}\label{e:SabMHV}
\tilde \cS_{4;a,b} &=& {1 \over 2^8} \left(S_{a,b}(12)^2 S_{a,b}(34)^2
- S_{a,b}(1234) -  S_{a,b}(1243)- S_{a,b}(1423)\right)\nn\\
 \tilde\cS_{2;a,b} &=&{1 \over 2^4}  (\partial \cP(12)-\partial
\cP(14))(\partial \cP(21)-\partial \cP(24))
(S_{a,b}(34))^2\\
\nn&+&{1 \over 2^4} (\partial  \cP(32)-\partial \cP(34))(\partial \cP(42)-\partial \cP(43))
(S_{a,b}(12))^2\,,\nn
\end{eqnarray}
where we have used a shorthand notation; $S_{a,b}(ij)$ stands for
$S_{a,b}(z_i-z_j)$ while $S_{a,b}(ijkl)$ stands for
$S_{a,b}(z_i-z_j)S_{a,b}(z_j-z_k) S_{a,b}(z_k-z_l)
S_{a,b}(z_l-z_i)$. 
With that choice of helicity, we can immediately give a simplified expression
for the contribution of a spin-$1$ supermultiplet in the $(4,0)$ models. We
introduce the field theory limit of $\tilde \cW^B$:
\begin{equation}
  \label{e:WBftdef}
  W^B:=\lim_{\tau_2 \to \infty} \left(\frac 2 \pi\right)^4 
\int_{-\frac12}^{\frac12}
\prod_{i=1}^4 \, d\nu^1_i \,
\tilde\cW^B|_{\bar q^0}\,.
\end{equation}
In this limit, this quantity is given by $W^B=W_1+ W_2$ with
\begin{eqnarray}
 \nn W_1&=& \, 
    {1\over16}(\partial P(12)-\partial P(14))
    (\partial P(21)-\partial P(24))
    (\partial P(32)-\partial P(34))
    (\partial P(42)-\partial P(43))\,,\\
 W_2&=& {1 \over 4\pi}{1\over\alpha'\tau_2 u}\;
\partial^2 P(24)
 (\partial P(12)-\partial P(14))
 (\partial P(32)-\partial P(34))\,,
\label{e:Wbft}
\end{eqnarray}
where $\partial^n P(\omega)$ is the $n$-th derivative of the field theory
propagator~\eqref{e:dPft} and where $\alpha'\tau_2$ is the proper time of
the field one-loop amplitude.
We can now rewrite \eqref{e:Ahetspin1} and find
\begin{equation}
   A^{spin-1}_{(4,0)het}=\frac14\,\left(\pi\over2\right)^8\,t_8t_8R^4\,
W^B \,.
\label{e:4ghetftspin1}
\end{equation}

Let us come back to the comparison of the $\cN=4$ spin-2
multiplet contributions in the type~II asymmetric orbifold
model given in~\eqref{e:IIspin2} and the heterotic one given
in~\eqref{e:Ahetspin2}.

We consider the following part of~\eqref{e:Ahetspin2}
\begin{equation}
 \int_{-\frac12}^{\frac12} \prod_{i=1}^4 \, d\nu^1_i \,
(\bar\cW^B|_{\bar q} (1+\alpha'\delta Q) + \bar\cW^B|_{\bar q^0}\cQ|_{\bar
q})\,,
\end{equation}
defined in the field theory limit for large $\tau_2$.

 The integral over the $\nu^1_i$ will kill any
term that have a non zero phase $e^{i \pi (a \nu_1^1 + b \nu_2^1 + c \nu_3^1 + d
\nu_4^1)}$ where $a,b,c,d$ are non-all-vanishing integers. 
In $\tilde W^B_1$ we have terms of the
form $\partial\delta P(ij) \times \partial \cP(ji)|_{\bar q} \times (\partial
P(kl)- \partial P(k'l')) (\partial
P(rs)- \partial P(r's'))$. Using the definition of $\delta\cP(ij)$ given
in~\eqref{e:delP} and the order $\bar q$ coefficient of the propagator 
 in~\eqref{e:Pg1qw}, we find that
\begin{equation}
\label{e:dP2}
 \partial\delta P(ij) \times \partial \cP(ji)|_{\bar q} = -{i
   \pi^2\over2}\,  \sin(2 \pi
\nu_{ij}) \,\sg(\omega_{ij})e^{2 i \pi
\sg(\omega_{ij}) \nu_{ij}}\,,
\end{equation}
which  integrates to $-\pi^2/4$. All such terms with $(ij)=(12)$ and $(ij)=(34)$
contribute  in total to
\begin{equation}
  \frac 12 \left({ \pi \over 2}\right)^4\left[(\partial P(12)-\partial P(14))
(\partial P(21)-\partial P(24)) 
+(\partial P(32)-\partial P(34))    (\partial P(42)-\partial P(43))\right]\,,
\end{equation}
where $\partial P(ij)$ is for the derivative of the propagator in
the field theory $\partial P(\omega_i-\omega_j)$
given by
\begin{equation}
  \label{e:dPft}
  \partial P(\omega)=2\omega-\sg(\omega)\,.
\end{equation}
The last contraction in $\tilde W^B_1$ for $(ij)=(24)$ 
leads to the same kind of contribution. However, they will actually be cancelled
by terms coming from similar 
contractions in $\tilde\cW^B_2|_{\bar q}$.
More precisely, the non zero contractions involved yield
\begin{equation}
 (\partial^2 \cP(24) e^\cQ )|_{\bar q} = -{\alpha'\pi^2\over2}\left(\cos(2 \pi
\nu_{24}) e^{2 i \pi \sg(\omega_{24})\nu_{24}} - 2 e^{2 i \pi
\sg(\omega_{24})\nu_{24}} \sin^2(\pi \nu_{24})\right)\,,
\end{equation}
which integrates to $-\alpha' \pi^2 /2$. The $\alpha'$ compensates the
$1/\alpha'$ factor in~\eqref{e:Wg} and this contribution precisely
cancels the one from~\eqref{e:dP2} with $(ij)=(24)$. Other types of
terms with more phase factors from the propagator turn
out to vanish after summation.
In all, we get $-\pi^4 W_3/4 $, where
\begin{equation}
  \label{e:W3ft}
W_3= -{ 1 \over 8}\,  \Big((\partial
P(12)-\partial P(14))
    (\partial P(21)-\partial P(24))+
    (\partial P(32)-\partial P(34))
    (\partial P(42)-\partial P(43))\Big)\, ,
\end{equation}
Finally, let us look at the totally contracted terms of the form
$\partial\delta P(ik) \partial\delta P(kl) \partial\delta P(lj) \times
\partial \cP(ij)|_{\bar q}$ that come from $\tilde\cW^B_1|{\bar q}$. Those are the only
terms of that type that survive the $\nu^1$ integrations
since they form closed chains in their arguments.
They give the following terms
\begin{equation}
i {\pi^4\over8} \sin (\pi \nu_{ij}) 
\sg(\omega_{ik})\sg(\omega_{kl})\sg(\omega_{lj})
e^{2 i \pi \left( \sg(\omega_{ik}) \nu_{ik}+\sg(\omega_{kl})
\nu_{kl}+\sg(\omega_{lj}) \nu_{lj}\right)}\,.
\end{equation}
They integrate to $ \pi^4/16$
if the vertex operators are ordered according to
$0\leq\omega_i<\omega_k<\omega_l<\omega_j\leq1$ or in the  reversed
ordering.
Hence, from $\tilde\cW^B_1$ we will get one of the orderings we want in our polarization
choice, namely the region $(s,t)$. From $\tilde\cW_2 e^\cQ$, a similar computation
yields the two other kinematical regions $(s,u)$ and $(t,u)$. In all we have a
total integrated contribution of  $\pi^4 /16$.
We collect all the different contributions that we have obtained, and
\eqref{e:Ahetspin2} writes:
\begin{equation}
  A^{spin-2}_{(4,0)het}=\frac14\,\left(\pi\over2\right)^8\,t_8t_8R^4\,
(1-4 W_3+2\, W^B) \,,
\label{e:4ghetft}
\end{equation}
where we used that $t_8t_8R^4= t_8F^4t_8\tilde F^4$ and
\eqref{e:WBftdef} and~\eqref{e:Wbft}.

We now turn to  the spin-2 contribution in the type~II asymmetric orbifold models
given in~\eqref{e:IIspin2}.
Using the $q$ expansion detailed in Appendix~\ref{sec:q-expansion}, we find
that 
\begin{equation}
  \int_{-\frac12}^{\frac12} \prod_{i=1}^4 \, d\nu^1_i \,
  (\tilde\cW_{0,0}|_{\sqrt q}-\tilde\cW_{0,1}|_{\sqrt q})= 
 \int_{-\frac12}^{\frac12} \prod_{i=1}^4 \, d\nu^1_i \,
  (\tilde\cW^F_{0,0}|_{\sqrt q}-\tilde\cW^F_{0,1}|_{\sqrt q})
=
2 \int_{-\frac12}^{\frac12} \prod_{i=1}^4 \, d\nu^1_i \,
\tilde\cW^F_{0,0}|_{\sqrt q} \,.
\end{equation}
We have then terms of the form $\tilde\cS_{2;0,0}$ and
$\tilde\cS_{4;0,0}$. Their structure is similar to the terms in the
heterotic case with, respectively, two and four 
bosonic propagators contracted.
The bosonic propagators do not have a $\sqrt q$ piece and since
$\tilde \cS_{0,0}(12)^2|_{\sqrt q}=\tilde \cS_{0,0}(34)^2|_{\sqrt q}=4\pi^2$ we find
that the terms in $\cS_{2;0,0}$ give
\begin{equation}
 2 \int_{-\frac12}^{\frac12} \prod_{i=1}^4 \, d\nu^1_i \,
\tilde\cS_{2;0,0}|_{\sqrt q}=-4\left({\pi \over2}\right)^4\, W_3\,,
\end{equation}
including the $1/2^4$ present in \eqref{e:Sdef}.
The $\tilde\cS_{4;0,0}$ terms have two different kind of contributions: double trace and
single trace (see, respectively, first and second lines in \eqref{e:Sdef}). In
the spin structure $(0,0)$ the double trace always vanishes in
$\tilde \cS_{4;0,0}|_{\sqrt q}$ since
\begin{equation}
\int_{-\frac12}^{\frac12} \prod_{i=1}^4 d\nu^1_i\, {\sin(\pi \nu_{ij}) \over \sin^2(\pi \nu_{kl})\sin(\pi \nu_{ij})} =\int_{-\frac12}^{\frac12} \prod_{i=1}^4 d\nu^1_i\,
{1\over \sin^2(\pi\nu_{kl}) }=0\,.
\end{equation}
However  the single trace terms are treated in the same
spirit as for the heterotic string. Only closed chains of sines contribute and are non zero
only for specific ordering of the vertex operators. For instance,
\begin{equation}
 -4\pi^4{\sin(\pi \nu_{ij}) \over \sin(\pi \nu_{jk})\sin(\pi \nu_{kl})\sin(\pi
\nu_{li})} \underset {\tau_2\to\infty}{\sim} -(2 \pi )^4\,,
\end{equation}
for the ordering $0\leq\omega_j<\omega_l<\omega_k<\omega_i\leq1$. Summing all of the contributions from $\tilde\cS_{4;0,0}$ gives a total factor of
$-\pi^4/16$, including the normalization in~\eqref{e:Sdef}.
We can now collect all the terms to get 
\begin{equation}
A^{spin-2}_{(4,0)II}=\frac14\,\left(\pi\over2\right)^8\,t_8t_8R^4\,
(1-4 W_3+2\,W^B) \,,
\label{e:4gIIft}
\end{equation}
showing the equality with the heterotic expression 
\begin{equation}
A^{spin-2}_{(4,0)het}= A^{spin-2}_{(4,0)II}\,.
\end{equation}
We remark that the same computations give the contribution of the spin-$\frac32$
multiplets in the two models, which are equal as well and write :

\begin{equation}
\label{e:4g32ft}
A^{spin-\frac32}_{(4,0)het}=A^{spin-\frac32}_{(4,0)II}=\frac14\,
\left(\pi\over2\right)^8\, t_8t_8R^4\, (W_3-2\,W^B) \,.
\end{equation}

Thanks to those equalities for the spin-$2$, spin-$\frac32$ and
spin-$1$ in~\eqref{e:40spin1eq}, from now we will use the notation
$A^{spin-s}_{(4,0)}$ with $s=1, \frac32, 2$.

\bigskip 
The perturbative equality between these two $(4,0)$ models is
not surprising. For a given number of vector multiplets $n_v$ the
heterotic and asymmetric type~II construction lead to two string
theory $(4,0)$ models related by $S$-duality, $S\to -1/S$, where $S$
is the axion-dilaton complex scalar in the $\cN=4$ supergravity
multiplet. The perturbative expansion 
in these two models is defined around 
different points in the  $SU(1,1)/U(1)$ moduli space. 
The action of $\cN=4$ supersymmetry implies that the one-loop amplitudes between
gravitons, which are neutral under the
$U(1)$ R-symmetry,  are the same in the strong and weak
coupling regimes.

\subsubsection{Comparing the $(4,0)$ and $(2,2)$ models}
\label{sec:comparing-4-0}

In the case of the $(2,2)$ models, the contribution  from the vector
multiplets is given in~\eqref{e:vm22bis}.
The string theory  integrand is different from the one in~\eqref{e:IIvm} for the
$(4,0)$ as it can be seen  using  the supersymmetric Riemann identity
in~\eqref{e:gsoym}. Let us first write the spin-$1$ contribution in the $(2,2)$
models. Performing the $\nu^1_i$ integrations and the same kind of manipulations
that we have done in the previous section, we can show that it is
given by
\begin{equation}
  \label{e:Aspin1diff22}
  A^{spin-1}_{(2,2)}=\frac14\,\left(\pi\over 2\right)^8\, t_8t_8R^4\,
((W_3)^2+\frac12\,W_2)\,.
\end{equation}
This is to be compared with \eqref{e:4ghetftspin1}. The expressions are clearly
different, but will lead to the same amplitude. In the
same manner, we find for the spin-$\frac32$:
\begin{equation}
  \label{e:Aspin32diff22}
  A^{spin-\frac32}_{(2,2)}=\frac14\,\left(\pi\over 2\right)^8\, t_8t_8R^4\,
(W_3-2((W_3)^2+\frac12\,W_2))\,.
\end{equation}
This differs from \eqref{e:4g32ft} by a factor coming solely from the vector
multiplets.

We now compare the spin-$2$ contributions in the $(4,0)$ model
in~\eqref{e:Amp32} and the $(2,2)$ model in~\eqref{e:22spin32}. Again, a
similar computation to the one we have done gives the contribution of the
spin-$2$ multiplet running in the loop for the $(2,2)$ model:
\begin{equation}
  \label{e:Aspin2diff22}
  A^{spin-2}_{(2,2)}=\frac14\,\left(\pi\over 2\right)^8\, t_8t_8R^4\,
  (1-4W_3+ 2((W_3)^2+\frac12\,W_2))\,.
\end{equation}
We compare this with \eqref{e:4ghetft} and \eqref{e:4gIIft} that we rewrite in
the following form:
\begin{equation}
  \label{e:Aspin2diff40}
A^{spin-2}_{(4,0)II} =\frac14\,\left(\pi\over 2\right)^8\, t_8t_8R^4\,
  (1-4W_3+ 2(W_1+W_2))\,.
\end{equation}
The difference between the two expressions originates again solely from the
vector multiplet sector. Considering that the same relation holds for the
contribution of the $\cN=4$ spin-$\frac32$ multiplets, we deduce that this is
coherent with the supersymmetric decomposition \eqref{e:multdec} that gives
\begin{equation}\label{e:M22final}
  A^{spin-2}_{(2,2)}= A^{spin-2}_{(4,0)} +2 \, (A^{spin-1}_{(2,2)}- A^{spin-1}_{(4,0)})\,.   
\end{equation}

The difference between the spin-2 amplitudes in the two models is
completely accounted for by the different vector multiplet contributions.
 The string theory models are related by a
U-duality exchanging the axion-dilaton scalar $S$ of the gravity
multiplet with a geometric modulus~\cite{Ferrara:1989nm,Witten:1995ex,Hull:1994ys}. 
This transformation affects the coupling of the multiplet running in
the loop, thus explaining the difference between the two string theory
models. 
However at the supergravity level, the four graviton amplitudes that we
compute are not sensitive to this fact and are equal in all models, as we will
see now.

\section{Field theory one-loop amplitudes in $\cN=4$ supergravity}
\label{sec:qftlimit-one-loop}

In this section we shall extract and compute the field theory limit $\alpha'\to
0$ of the one-loop string theory amplitudes studied in previous sections. We
show some relations between loop momentum power counting
and the spin or supersymmetry of the multiplet running in the loop.

As mentioned above,
the region of the fundamental domain integration corresponding to the
field theory amplitude is $\tau_2 \to \infty$, such that $t=\alpha'\,\tau_2$
is fixed. We then obtain a world-line integral of total proper time
$t$. The method for extracting one-loop field theory amplitudes from
string theory was pioneered in~\cite{Green:1982sw}. The general method that
we apply consists in extracting the $o(q)^0$ terms in the integrand and
taking the field theory limit and was developed extensively
in~\cite{Bern:1991aq,Bern:1993wt,Dunbar:1994bn}. Our approach will
follow the formulation given in~\cite{BjerrumBohr:2008ji}. 

The generic form of the field theory four-graviton one-loop amplitude
for $\cN=4$ supergravity with a spin-$s$ ($s=1, \frac32, 2$) $\cN=4$
supermultiplet running is the loop is given by 
\begin{equation}
  \label{e:1loopGeneric}
  M^{spin-s}_{X}=\left(4\over\pi\right)^4\,  {\mu^{2\epsilon}\over\pi^{D\over2}} \int_0^\infty {dt\over t^{D-6\over2}}\, \int_{\Delta_\omega}
  \prod_{i=1}^3 d\omega_i\,\, e^{-\pi\,t\,Q(\omega)}\times A_X^{spin-s}\,,
\end{equation}
where $D=4-2\epsilon$ and $X$ stands for the model, $X=(4,0)het$, $X=(4,0)II$ or
$X=(2,2)$ while the respective amplitudes $A^{spin-s}_X$ 
are given in sections~\ref{sec:ampl-contr}
and~\ref{sec:comp-string-models}. We have set the overall
normalization to unity.

The domain of integration $\Delta_\omega=[0,1]^3$ is decomposed into three regions
$\Delta_w=\Delta_{(s,t)}\cup \Delta_{(s,u)}\cup \Delta_{(t,u)} $  given by the union of
the $(s,t)$, $(s,u)$ and $(t,u)$ domains.
 In the $\Delta_{(s,t)}$ domain the integration
is performed over  $0\leq \omega_1\leq \omega_2\leq\omega_3\leq1$ where
$Q(\omega)=-s\omega_1(\omega_3-\omega_2)-t(\omega_2-\omega_1)(1-\omega_3)$ with equivalent
formulas obtained by permuting the external legs labels in the $(t,u)$ and $(s,u)$ regions
(see~\cite{Green:1999pv} for details). We used that $s=-(k_1+k_2)^2$, $t=-(k_1+k_4)^2$ and
$u=-(k_1+k_3)^2$ with our convention for the metric $(-+\cdots+)$.

We now turn to the evaluation  of the amplitudes.  The main properties of the bosonic and fermionic
propagators are provided in Appendix~\ref{sec:green-functions}.
We work with the helicity configuration detailed in the previous section.  This choice of
polarization makes the intermediate steps easier as the expressions
are explicitly gauge invariant.  

\subsection{Supersymmetry in the loop}
Before evaluating the amplitudes we discuss the action of supersymmetry
on the structure of the one-loop amplitudes. An $n$-graviton amplitude in
dimensional regularization with $D=4-2\epsilon$ can generically be written in the
following way:
\begin{equation}\label{e:Mn1Fey}
  M_{n;1}= \mu^{2\epsilon}\,   \int {d^D\ell\over(2\pi)^D}\, {\mathfrak
    N(\epsilon_i,k_i;\ell)\over \ell^2 (\ell-k_1)^2 \cdots
    (\ell-\sum_{i=1}^{n-1} k_i)^2}\,,
\end{equation}
where the numerator is a
polynomial in the loop momentum $\ell$ with coefficients depending on
the external momenta $k_i$  and polarization of the gravitons $\epsilon_i$.  For $\ell$
large this numerator behaves as $\mathfrak N(\epsilon_i,k_i;\ell)\sim
\ell^{2n}$ in non-supersymmetric theories. In an $\cN$ extended
supergravity theory, supersymmetric cancellations improve this
behaviour, which becomes $\ell^{2n-\cN}$ where $\cN$ is the number of
four-dimensional supercharges:
\begin{equation}\label{e:Nasymp}
  \mathfrak N^{\cN}(\epsilon_i,k_i;\ell)\sim    \ell^{2n-\cN}\,\qquad\textrm{for} \qquad |\ell|\to\infty\,.
\end{equation}
The dictionary between the Feynman integral presentation given
in~\eqref{e:Mn1Fey} and the structure of the field theory limit of the
string theory amplitude states that
the first derivative of a bosonic propagator counts as one power of loop
momentum $\partial \cP \sim \ell$, $\partial^2\cP\sim\ell^2$ while
fermionic propagators count for zero power of loop momentum
$S_{a,b}\sim 1$.
This dictionary was first established in~\cite{Bern:1991aq} for gauge
theory computation, and applied to
supergravity amplitudes computations in~\cite{Dunbar:1994bn}  and more
recently in~\cite{BjerrumBohr:2008ji}.

With this dictionary we find that in the $(4,0)$ model the integrand of
the amplitudes have the following behaviour
\begin{eqnarray}\label{e:Nasymp40}
  A^{spin-1}_{(4,0)}&\sim&\ell^4\,, \cr
  \label{e:loop40}
  A^{spin-\frac32}_{(4,0)}&\sim& \ell^2+\ell^4\,,\\
\nn  A^{spin-2}_{(4,0)}&\sim&1+\ell^2+\ell^4 \,.
\end{eqnarray}
The spin-1 contribution to the four-graviton amplitude has four powers
of loop momentum as required for an $\cN=4$ amplitude
according~\eqref{e:Nasymp}. The $\cN=4$ spin-$\frac32$ supermultiplet
contribution can be decomposed into an $\cN=6$ spin-$\frac32$
supermultiplet term with two powers of loop momentum, and an $\cN=4$
spin-1 supermultiplet contribution with four powers of loop
momentum. The spin-2 contribution has an $\cN=8$ spin-2 piece with no
powers of loop momentum, an $\cN=6$ spin-$\frac32$ piece with two
powers of loop momentum and an $\cN=4$ spin-1 piece with four powers
of loop momentum.

For the $(2,2)$ construction we have the following behaviour 
\begin{eqnarray}\label{e:Nasymp22}
  A^{spin-1}_{(2,2)}&\sim&(\ell^2)^2 \,,\cr
  \label{e:loop22} A^{spin-\frac32}_{(2,2)}&\sim& \ell^2+(\ell^2)^2\,,\\
\nn  A^{spin-2}_{(2,2)}&\sim&1+\ell^2+(\ell^2)^2 \,.
\end{eqnarray}
Although the superficial counting of the number of loop momenta is the
same for each spin-$s=1,\frac32, 2$ in the two
models, the precise dependence on the loop momentum  differs in the
two models, as indicated 
by  the symbolic notation $\ell^4$ and $(\ell^2)^2$. 
This is a manifestation of the model dependence for the vector multiplet contributions.
As we have seen in the previous section, the order four terms in the loop
momentum in the
spin-$\frac32$ and spin-$2$ parts are due to the spin-1 part.

At the level of the string amplitude, the multiplets running in the loop
(spin-2 and spin-1) are naturally decomposed under the $\cN=4$ supersymmetry
group. However, at the level of the amplitudes in field theory it is convenient
to group the various blocks according to the number of powers of loop momentum
in the numerator
\begin{equation}
  A^{spin-s}_{\cN=4s} \sim \ell^{4(2-s)},  \qquad s=1, \frac32, 2\,,  
\end{equation}
 which is the same as organizing the terms according to
the supersymmetry of the corresponding $\cN=4s$ spin-$s=1, \frac32, 2$
supermultiplet. In this decomposition it is understood that for the two $\cN=4$
models the dependence in the loop momenta is not identical.

From these blocks, one
can reconstruct the contribution of the spin-2 $\cN=4$ multiplet that we are
concerned with using the following relations
\begin{equation}
  M_{X}^{spin-2}=M_{\cN=8}^{spin-2}-4
M^{spin-\frac32}_{\cN=6}+2 M_{X}^{spin-1}\,,
\label{e:ampdecomp}
\end{equation}
where the index $X$ refers to the type of model, $(4,0)$ or $(2,2)$.

This supersymmetric decomposition 
of the one-loop amplitude reproduces the one given
in~\cite{Dunbar:1994bn,Dunbar:1995ed,Dunbar:2010fy,Dunbar:2011xw,Dunbar:2011dw,
Dunbar:2012aj,Bern:2011rj}.

We shall come now to the evaluation of those integrals. We will see that even
though the spin-$1$ amplitudes have different integrands, \ie~ different loop
momentum dependence in the numerator of the Feynman integrals, they are equal
after integration.

\subsection{Model-dependent part : $\cN=4$  vector 
 multiplet contribution}\label{sec:QFTvm}

In this section we first compute the field theory amplitude with an $\cN=4$
vector multiplet running in the loop for the two models. This part of the
amplitude is model dependent as far as concerns the
integrands. However, the value of the
integrals is the same in the different models. 
Then we provide an analysis of the IR and UV behaviour of these amplitudes.

\subsubsection{Evaluation of the field theory amplitude}
\label{sec:eval-field theory}

The contribution from the $\cN=4$ spin-1 vector supermultiplets in the $(4,0)$
models is %
\begin{equation}\label{e:AmpN4hetvm}
  M^{spin-1}_{(4,0)}=\left(4\over\pi\right)^4\,
{\mu^{2\epsilon}\over\pi^{D\over2}}\, \int_0^\infty {dt\over t^{D-6\over2}}\,
\int_{\Delta_\omega}
 d^3\omega\,\, e^{-\pi\,t\,Q(\omega)}\times A^{spin-1}_{(4,0)}\,,
\end{equation}
where $A^{spin-1}_{(4,0)}$ is given in~\eqref{e:4ghetftspin1} for instance and
$Q$
defined in~\eqref{e:Qft}.
Integrating over the proper time $t$ and setting $D=4-2\epsilon$, the amplitude
reads 
\begin{equation}\label{e:N=4matter}
    M^{spin-1}_{(4,0)} =t_8t_8R^4\,
  \int_{\Delta_\omega} \!\!\!d^3\omega\!\left[ \Gamma\left(1+\epsilon\right)\,
  Q^{-1-\epsilon}\, W_2
+\Gamma\left(2+\epsilon\right)\, Q^{-2-\epsilon} \, W_1\,\right]\,.
\end{equation}
The quantities $W_1$ and $W_2$ are given in~\eqref{e:Wbft}, they have the following form in
terms of the variables $\omega_i$:
\begin{eqnarray}
 \nn W_1&=&{1\over8}\,
( \omega_2- \omega_3) (\sg(\omega_1-\omega_2)+2 \omega_2-1)
   (\sg(\omega_2-\omega_1)+2 \omega_1-1) (\sg(\omega_3-\omega_2)+2
   \omega_2-1)\\
 \label{e:Tneval}
W_2&=&-{1\over4}\;{1\over u}\, (2\omega_2-1+\sg(\omega_3-\omega_2)) 
(2\omega_2-1+\sg(\omega_1-\omega_2))\,(1-\delta(\omega_{24}))
\,.
\end{eqnarray}
Using the dictionary between the world-line propagators and  the
Feynman integral from the string-based rules~\cite{Bern:1991aq,Dunbar:1994bn,BjerrumBohr:2008ji},
 we recognize in  the
first term in~\eqref{e:N=4matter}  a six-dimensional scalar box
integral and in
the second term four-dimensional scalar bubble
integrals.\footnote{In~\cite{Bern:2007xj,BjerrumBohr:2008ji} it
  was wrongly claimed that $\cN=4$ amplitudes do not have rational
  pieces. The argument in~\cite{BjerrumBohr:2008ji} was based on a naive application of the reduction formulas
  for $\cN=8$ supergravity amplitudes to $\cN=4$ amplitudes where boundary
  terms do not cancel anymore.}
Evaluating the integrals with standard techniques, we find\footnote{The
analytic continuation in the complex energy plane corresponds to the
$+i\varepsilon$ prescription for the Feynman propagators
$1/(\ell^2-m^2+i\varepsilon)$. We are using the notation that $\log(-s)=
\log(-s-i\varepsilon)$ and that
$\log(-s/-t):=\log((-s-i\varepsilon)/(-t-i\varepsilon))$. }
\begin{equation}\label{e:VM40}
M^{spin-1}_{(4,0)} =  {t_8 t_8 R^4 \over 2s^4} \left( s^2 -
s(u-t)\log\left(-t\over-u\right) -tu
(\log^2\left(-t\over-u\right)+\pi^2)
\right)\,.
\end{equation}
The crossing symmetry of the amplitude  has been broken by our choice
of helicity configuration. However, it is still invariant under the exchange of
the legs $1\leftrightarrow2$ and $3\leftrightarrow4$ which amounts to exchanging
$t$ and $u$. The same comment applies to all the field
theory amplitudes evaluated in this paper.
This result matches the one derived
in~\cite{Dunbar:1994bn,Dunbar:1995ed,Dunbar:2010fy,Dunbar:2011xw,Dunbar:2011dw,Dunbar:2012aj}
and in particular~\cite[eq.~(3.20)]{Bern:2011rj}.

Now we turn to the amplitude in the $(2,2)$ models:
\begin{equation}\label{e:AmpN4II}
  M^{spin-1}_{(2,2)}=\left(4\over\pi\right)^4\,
{\mu^{2\epsilon}\over\pi^{D\over2}} \int_0^\infty {dt\over t^{D-6\over2}}\,
\int_{\Delta_\omega}
d^3\omega\,\, e^{-\pi\,t\,Q(\omega)}\times A^{spin-1}_{(2,2)}\,,
\end{equation}
where $A^{spin-1}_{(2,2)}$ is defined in~\eqref{e:vm22bis}.
After integrating over the proper time $t$, one gets
\begin{equation}
 \label{e:AvmII22}    
M^{spin-1}_{(2,2)}={t_8t_8R^4}\,
  \int_{\Delta_\omega} \!\!\!d^3\omega\,
[\Gamma\left(2+\epsilon\right)\, Q^{-2-\epsilon} \,(W_3)^2+\frac 12
\Gamma(1+\epsilon) \, Q^{-1-\epsilon}\, W_2]\,,
\end{equation}
where $W_3$ defined in~\eqref{e:W3ft}, is given  in terms
of the $\omega_i$ variables by 
\begin{multline}\label{e:W3}
W_3=-\frac18\,
(\sg(\omega_1-\omega_2)+2 \omega_2-1)(\sg(\omega_2-\omega_1)+2 \omega_1-1)\cr
+\frac14\, (\sg(\omega_3-\omega_2)+2\omega_2-1)( \omega_3- \omega_2) \,.
\end{multline}
There is no obvious relation between the integrand of this amplitude
with the one for $(4,0)$ model in~\eqref{e:N=4matter}.
Expanding the square one can decompose this integral in three pieces that are seen to be
proportional to the $(4,0)$ vector multiplet contribution
in~\eqref{e:VM40}. A first contribution is
\begin{equation}
    {t_8t_8R^4\over2}\,
  \int_{\Delta_\omega} \!\!\!d^3\omega\,
[\Gamma\left(2+\epsilon\right)\, Q^{-2-\epsilon} \,W_1+
\Gamma(1+\epsilon) \, Q^{-1-\epsilon}\, W_2]  = \frac 12 \, M_{(4,0)}^{spin-1}
\end{equation}
and we have the additional contributions
\begin{equation}
      {t_8t_8R^4\over 64}\,
  \int_{\Delta_\omega} \!\!\!d^3\omega\,
{\Gamma\left(2+\epsilon\right)\over Q^{2+\epsilon}} \,
((\sg(\omega_1-\omega_2)+2 \omega_2-1)(\sg(\omega_2-\omega_1)+2 \omega_1-1))^2 
= \frac 14 \, M_{(4,0)}^{spin-1}
\end{equation}
and
\begin{equation}
  {t_8t_8R^4\over64}\,
  \int_{\Delta_\omega} \!\!\!d^3\omega\,
{\Gamma\left(2+\epsilon\right)\over Q^{2+\epsilon}} \,(
(\sg(\omega_3-\omega_2)+2\omega_2-1)( \omega_3- \omega_2))^2  = \frac 14 \,
M_{(4,0)}^{spin-1}\,. \end{equation}

Performing all the integrations leads to 
\begin{equation}\label{e:VM22}
M^{spin-1}_{(2,2)} = M_{(4,0)}^{spin-1}\,.
\end{equation}

It is now clear that the vector multiplet contributions to the amplitude are
equal
in the two theories, $(4,0)$ and $(2,2)$. It would be interesting to see if this
expression could be derived with the double-copy construction
of~\cite{Bern:2011rj}. 

 In this one-loop 
amplitude there is no interaction between the vector
multiplets. Since the coupling of individual vector multiplet to
gravity  is universal (see for instance the $\cN=4$ Lagrangian
given in~\cite[eq.(4.18)]{de Roo:1984gd}), the four-graviton one-loop
amplitude in pure $\cN=4$ supergravity has to be independent of the model
it comes from.

\subsubsection{IR and UV behaviour}
\label{sec:ir-uv-behaviour}

\begin{figure}[t]
  \centering
  \includegraphics[width=15cm]{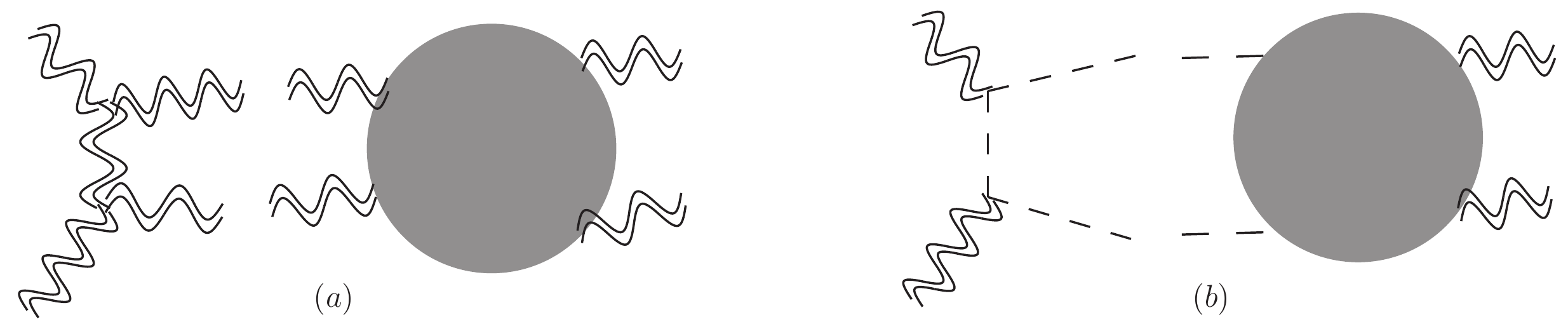}
  \caption{Contribution to the IR divergences when two external gravitons (double
wavy lines) become soft. If a graviton is exchanged as in (a) the amplitude
presents an IR divergence. No IR divergences are found when another massless
state of spin different from two is exchanged as in~(b).}
  \label{fig:softgrav}
\end{figure}

The graviton amplitudes with vector multiplets running in the
loop in~\eqref{e:VM40} and~\eqref{e:VM22} are free of UV and IR divergences.  
The absence of IR divergence is expected, since no spin-2 state is running in
the loop. The IR divergence occurs only when a graviton is exchanged between two
soft graviton legs (see figure~\ref{fig:softgrav}).  This fact has already
been noticed in~\cite{Dunbar:1995ed}. 

\bigskip
This behaviour is easily understood  by considering the soft graviton
limit of the coupling between the graviton and a spin-$s\neq2$ state.
It occurs through the stress-energy tensor $V^{\mu\nu}(k,p)=T^{\mu\nu}(p-k,p)$
where $k$ and $p$ are, respectively, the momentum of the graviton
and of the exchanged state.
In the soft graviton  limit the vertex  behaves as 
$V^{\mu\nu}(p-k,p)\sim -k^\mu p^\nu$ for $p^\mu\sim 0$, and the amplitude
behaves in the soft limit as
\begin{equation}
  \int_{\ell\sim 0} {d^4\ell\over \ell^2 (\ell\cdot k_1)(\ell  \cdot k_2)}\,
  T_{\mu\nu}(\ell-k_1,\ell)  T^{\mu\nu}(\ell,\ell+k_2)\sim  (k_1\cdot k_2) 
\int_{\ell\sim_0} {d^4\ell\over \ell^2 (\ell\cdot k_1)(\ell  \cdot k_2)}\,
 \ell^2 \,,
\end{equation}
which is finite for small values of the loop momentum $\ell\sim 0$.
In the soft graviton limit, the three-graviton vertex behaves as $V^{\mu\nu}(k,p)\sim
k^\mu k^\nu$ and the amplitude has a logarithmic divergence at
$\ell\sim 0$:
\begin{equation}
  (k_1\cdot k_2)^2 \int_{\ell\sim 0} {d^4\ell\over \ell^2 (\ell\cdot k_1)(\ell 
\cdot k_2)}=\infty \,.
\end{equation}

The absence of UV divergence is due to the fact that the $R^2$
one-loop counter-term  is the Gauss-Bonnet term. It vanishes in the
four-point amplitude since it is a total derivative~\cite{'tHooft:1974bx}.

\subsection{Model-independent part}\label{sec:modindep}

In this section we compute the field theory amplitudes with an $\cN=8$
supergraviton and an  $\cN=6$ spin-$\frac32$ supermultiplet 
running in the loop. These quantities are model independent in the sense that
their integrands are the same in the different models.

\subsubsection{The $\cN=6$ spin-$\frac32$ supermultiplet contribution}\label{sec:spin32}

The integrand for the $\cN=4$ spin-$\frac32$ supermultiplet contribution is different in the two
$(4,0)$ and $(2,2)$ constructions of the $\cN=4$ supergravity
models. As shown in equations~\eqref{e:4g32ft} and.~\eqref{e:Aspin32diff22},  
this is accounted for by the contribution of the vector
multiplets. However, we  exhibit an $\cN=6$ spin-$\frac32$
supermultiplet model-independent piece by adding  two  $\cN=4$  vector
multiplet  contributions to the one of an $\cN=4$ spin-$\frac32$ supermultiplet 
\begin{equation}
   M^{spin-\frac32}_{\cN=6} =M^{spin-\frac32}_{X}+2M^{spin-1}_{X}\,.
\end{equation}
The amplitude with an $\cN=6$ spin-$\frac32$ multiplet running in the
loop is
\begin{equation}\label{e:N=4Gravitino1loop}
  M^{spin-\frac32}_{\cN=6} = -{t_8 t_8 R^4\over8} \,
  \int_{\Delta_\omega}
  d^3\omega\,\Gamma\left(2+\epsilon\right)\, W_3\, Q^{-2-\epsilon},
\end{equation}
where $W_3$ is given in~\eqref{e:W3}.
The integral is
equal  to the six-dimensional scalar box
integral given in~\cite[eq.~(3.16)]{Bern:2011rj} up to $o(\epsilon)$
terms. We evaluate it, and get
\begin{equation}\label{e:N3half}
M^{spin-\frac32}_{\cN=6} =- {t_8 t_8 R^4 \over2 s^2}\,
\left(\log^2\left(-t\over-u\right)+\pi^2 \right)\,.
\end{equation}
This result is UV finite as  expected from the superficial power counting of
loop momentum in the numerator of the amplitude given
in~\eqref{e:Nasymp40}. It is free of IR
divergences because no graviton state is
running in the loop   (see the previous section).  It
matches the one derived 
in~\cite{Dunbar:1994bn,Dunbar:1995ed,Dunbar:2010fy,Dunbar:2011xw,Dunbar:2011dw,Dunbar:2012aj}
and in particular~\cite[eq.~(3.17)]{Bern:2011rj}.

\subsubsection{The $\cN=8$ spin-2 supermultiplet contribution}

We now turn to the $\cN=8$ spin-2 supermultiplet contribution in~\eqref{e:ampdecomp}. 
 It has already been evaluated
 in~\cite{Grisaru:1981ye,Green:1982sw} and can be written as:
\begin{equation}\label{e:N=8}
    M^{spin-2}_{\cN=8} = {t_8t_8R^4\over4}\, 
  \int_{\Delta_\omega} d^3\omega\,\Gamma\left(2+\epsilon\right)\,
Q^{-2-\epsilon}\,.
\end{equation}
Performing the integrations we have 
\begin{eqnarray}
\label{e:N8} M^{spin-2}_{\cN=8}&=&{t_8 t_8 R^4\over4}\,\left[{2 \over \epsilon} \left(
{\log\left(-t\over\mu^2\right) \over s
u}+{\log\left(-s\over\mu^2\right) \over t u}+{\log\left(-u\over\mu^2\right)
\over
s t} \right)\right. + \\
\nn&+&\left.2 \left( {\log\left(-s\over\mu^2\right)
\log\left(-t\over\mu^2\right) \over st }+ {\log\left(-t\over\mu^2\right)
\log\left(-u\over\mu^2\right) \over tu
}+ {\log\left(-u\over\mu^2\right) \log\left(-s\over\mu^2\right) \over
us}\right)\right]\,,
\end{eqnarray}
where $\mu^2$ is an IR mass scale. This amplitudes carries an $\epsilon$ pole
signaling the IR divergence due to the graviton running in the loop.

Now we have all the blocks entering the expression for the $\cN=4$
pure gravity amplitude in~\eqref{e:ampdecomp}.

\section{Conclusion}\label{sec:conclusion}

In this work we have evaluated the four-graviton amplitude at one loop
in $\cN=4$ supergravity in four dimensions from the field theory limit of string
theory constructions. The string theory approach includes $(4,0)$
models where all of the supersymmetry come from the left-moving sector of
the theory, and $(2,2)$ models  where the supersymmetry is split between
the left- and right-moving sectors of the theory.

For each model the four-graviton one-loop amplitude is linearly 
dependent on the number of vector multiplets $n_v$. Thus
we define the  pure  $\cN=4$ supergravity amplitude by subtraction of 
these contributions. This matches the result obtained in
the Dabholkar-Harvey construction of string theory models with no
vector multiplets.
We have seen that, except when gravitons are running
in the loop, the one-loop amplitudes are free of IR divergences. In addition,
all the amplitudes are UV finite because the $R^2$ candidate counter-term
vanishes for these amplitudes. Amplitudes with external vector states
are expected to be UV divergent~\cite{Fischler:1979yk}.

Our results reproduce  the ones obtained with the
string-based rules in~\cite{Dunbar:1994bn,Dunbar:1995ed}
unitarity-based method
in~\cite{Dunbar:2010fy,Dunbar:2011xw,Dunbar:2011dw,Dunbar:2012aj}
and the double-copy approach of~\cite{Bern:2011rj}. 
The structure of the string theory amplitudes of the $(4,0)$ and  $(2,2)$ models
takes a very different form. There could have been  differences at the
supergravity level due to the different nature of the couplings of the
vector multiplet in the two theories as indicated by the relation
between the two amplitudes in~\eqref{e:M22final}. However, the coupling to
gravity is universal. The difference
between the various $\cN=4$ supergravity models are visible once interactions
between vectors and scalars occur, as can be seen on structure of the $\cN=4$
Lagrangian in~\cite{de Roo:1984gd}, which is not the case in our amplitudes
since they involve only external gravitons. Our computation provides a direct
check of this fact.

The supergravity amplitudes studied in this paper are 
naturally organized as a sum 
of $\cN=4s$ spin-$s=1, \frac32, 2$ contributions, with a
simple power counting dependence on the loop momentum
$\ell^{4(2-s)}$. Such a decomposition has been already used in the
string-based approach to supergravity amplitudes in~\cite{Dunbar:1994bn}. Our
analysis reproduces these results and shows that the $\cN=4$ part of the
four-graviton amplitude does not depend on
whether one starts from $(4,0)$ or $(2,2)$ construction.  We expect amplitudes
with external scalars or vectors to take a different form in the two
constructions.

\section*{Acknowledgements}

We would like to thank  Ashoke Sen for discussions and in particular Arniban Basu for
having pointed out to us the construction by Dabholkar and Harvey
in~\cite{Dabholkar:1998kv}. We would like to thank Guillaume Bossard, Renata Kallosh and
Warren Siegel for comments on a previous version of this paper. We
particularly thank Nick Halmagyi for comments on the manuscript.

\newpage
\appendix

\section{World-sheet CFT : chiral blocks, propagators.}
\label{sec:chiralblocks}

In this Appendix we collect various results about the conformal blocks,
fermionic and bosonic propagators at genus one, and their $q$ expansions.

\subsection{Bosonic and fermionic chiral blocks}

\noindent$\triangleright$
The genus one theta functions are defined to be
\begin{equation}
  \theta\left[a\atop b\right](z|\tau)=\sum_{n\in\ZZ}
  q^{{1\over2}(n+\frac a2)^2} \,
  e^{2i\pi (z+\frac b2)(n+\frac a2)}\,,
\end{equation}
and Dedekind eta function:
\begin{equation}
  \eta(\tau)= q^{1\over24}\prod_{n=1}^\infty (1-q^n)\,,
\end{equation}
where $q=\exp(2i\pi \tau)$. Those functions have the following $q\to0$
behaviour:
\begin{eqnarray}
\theta\left[1\atop 1\right](0,\tau)&=&0\,;\quad \theta\left[1\atop
0\right](0,\tau)= -2 q^{1/8}+o(q)\,;\quad\theta\left[0\atop
0\right](0,\tau)=1+2\sqrt q+o(q)\,;\nn\\
 \theta\left[0\atop 1\right](0,\tau)&=&1-2\sqrt q+o(q)\,;\quad
\eta(\tau) =q^{1/24}+o(q)\,.
\label{e:thetaq}
\end{eqnarray}

\noindent$\triangleright$ The partition function of eight world-sheet
fermions  in the $(a,b)$-spin structure, $\Psi(z+1)=-(-1)^{2a}\Psi(z)$
and $\Psi(z+\tau)=-(-1)^{2b}\Psi(z)$, 
 and eight chiral bosons is
\begin{equation}\label{e:ZpsiEven}
Z_{a,b}(\tau)\equiv {\theta\left[a\atop b\right](0|\tau)^4\over \eta^{12}(\tau)}\,,
\end{equation}
it has the following behaviour for $q\to 0$ 
\begin{eqnarray}
Z_{1,1}&=&0\,,\cr
Z_{1,0}&=&16+16^2 q+o(q^2)\,,\cr
\label{e:Zasymp} Z_{0,0}&=&{1\over \sqrt q}+8+o(\sqrt q)\,,\\
\nn Z_{0,1}&=&{1\over \sqrt q}-8+o(\sqrt q)\,.
\label{e:chblksexp}
\end{eqnarray}

\noindent$\triangleright$  The partition function of the twisted
$(X,\Psi)$ system in the  $(a,b)$-spin structure is

\begin{eqnarray}
  X(z+1)&=(-1)^{2h} \, X(z);\qquad \Psi(z+1)&=-(-1)^{2a+2h} \Psi(z)\,,\cr
X(z+\tau)&=(-1)^{2g}\,X(z);\qquad \Psi(z+\tau)&=-(-1)^{2b+2g}\,\Psi(z)\,.    
\end{eqnarray}
The twisted chiral blocks for a real boson are
\begin{equation}
 \cZ^{h,g}[X]=\left(i e^{-i\pi g} q^{-h^2/2} 
{\eta(\tau) \over\theta\left[ 1+h
\atop 1+g \right] }\right)^{1/2}\,.
\label{e:chblkbos}
\end{equation}
The  twisted chiral blocks for a Majorana or Weyl fermion are
\begin{equation}
 \cZ^{h,g}_{a,b}[\Psi]=\left(e^{-i\pi\, a(g+b)/2} q^{h^2/2} 
{ \theta\left[ a+h
\atop b+g \right]\over \eta(\tau)}\right)^{1/2}\,.
\label{e:chblkferm}
\end{equation}
The total partition function is given by 
\begin{eqnarray}
  \cZ_{a,b}^{h,g}[(X,\Psi)]=\cZ^{h,g}[X]
  \cZ^{h,g}_{a,b}[\Psi]=e^{i{\pi\over4} (1+2g+a(g+b))}\, \sqrt{
   \theta\left[a+ h\atop b+g\right]\over
    \theta\left[1+h\atop 1+g\right]}\,.
\end{eqnarray}

\subsection{Bosonic and fermionic propagators}
\label{sec:green-functions}

\subsubsection{Bosonic propagators}
\label{sec:bosprop}

Our convention for the bosonic propagator is 

\begin{equation}
  \langle x^\mu( \nu) x^\nu(0)\rangle_{one-loop}= 2\alpha'\, \eta^{\mu\nu}\,\cP(\nu|\tau)\,,
\end{equation}
with
\begin{eqnarray}\label{e:newv} 
\cP(\nu|\tau)
&=&-{1\over 4} \ln\left|\theta\left[1\atop 1\right](\nu|\tau)\over \partial_\nu \theta\left[1\atop 1\right](0|\tau) \right|^2 +{\pi \nu_2^2\over 2\tau_2}+C(\tau)
\cr
&=&{\pi \nu_2^2\over 2\tau_2}- {1\over4} \ln\left|\sin(\pi \nu)\over \pi \right|^2 -
\sum_{m\geq1} \left({q^m\over 1-q^m} {\sin^2(m\pi \nu)\over m}
  +c.c.\right) +C(\tau),
\end{eqnarray}
where $C(\tau)$ is a contribution of the zero modes (see
e.g. \cite{Green:1999pv}) that anyway drops out of the string amplitude because of
momentum conservation so we will forget it in the following.

We have as well the expansions 
\begin{eqnarray}\label{e:dP}
  \partial_\nu \cP(\nu|\tau) &=&{\pi\over2i}\, {\nu_2\over\tau_2}-
  {\pi\over4} {1\over\tan(\pi \nu)}-\pi\,q\, \sin(2\pi\nu)+o(q) \,,\cr
  \partial_\nu^2 \cP(\nu|\tau) &=&-{\pi\over 4 \tau_2}
  +{\pi^2\over4}{1\over\sin^2(\pi\nu)}-2\pi^2 \,
  q\cos(2\pi\nu)+o(q)\cr
\partial_\nu\bar\partial_{\bar\nu}\cP(\nu|\tau)&=&{\pi\over4}\left({1\over\tau_2}-\delta^{(2)}(\nu)\right)\,.
\end{eqnarray}
leading to the following Fourier expansion with respect to $\nu_1$
\begin{eqnarray}
  \partial_\nu \cP(\nu|\tau) &=&{\pi\over4i}\, ({2\nu_2\over\tau_2}- \sg(\nu_2))
+i{\pi\over4}\,\sg(\nu_2)
  \,\sum_{m\neq0} e^{2i\pi m \sg(\nu_2) \nu}-\pi \,q\,
\sin(2\pi \nu)+o(q)\,,\nn\\
  \partial_\nu^2 \cP(\nu|\tau) &=&{\pi\over 4 \tau_2} \,(\tau_2\delta(\nu_2)-1)
-\pi^2\, \,\sum_{m\geq1} m e^{2i\pi m\sg(\nu_2)\nu}-2\pi^2
  \,q \,\cos(2\pi\nu)+o(q)\,.
\label{e:d2P}\end{eqnarray}
Setting  $\nu=\nu_1+i\tau_2\omega$ we can rewrite these expressions in
a form relevant for the field theory limit $\tau_2\to\infty$ with
$t=\alpha'\tau_2$ kept fixed.
 The bosonic propagator can be decomposed in an asymptotic value for
$\tau_{2}\to\infty$
(the field theory limit) and corrections originating from massive string modes

\begin{equation}\label{e:Pg1qw} 
\cP(\nu|\tau)=-\,{ \pi \,t \over2\alpha'}\,  P(\omega) + \delta P(\nu)-q\,
\sin^2(\pi \nu)-\bar q\,
\sin^2(\pi \bar \nu)+o(q^2)\,,
\end{equation}
and
\begin{equation}
  \label{e:delP} 
P(\omega) = \omega^2 - |\omega|;\qquad
  \delta P(\nu)= \sum_{m\neq0}{1\over 4|m|}\, e^{2i\pi m\nu_1-2\pi |m\nu_2|}\,.
\end{equation}
The contribution $\delta P$ corresponds to the effect of massive string
states propagating between two external massless states.
 The quantity $\cQ$ defined in~\eqref{e:QstringDef} writes in this limit 
\begin{equation}
 \mathcal Q=-t \pi Q(\omega) +\alpha' \delta
 Q-2\pi\alpha'\,\sum_{1\leq i<j\leq 4} k_i\cdot
 k_j \,(q \sin^2(\pi \nu_{ij})+\bar q\sin^2(\pi\bar\nu_{ij}))  +o(q^2)\,,
\label{e:delQ}
\end{equation}
where
\begin{equation}
  \label{e:Qft}
  Q(\omega)=\sum_{1\leq i<j\leq 4} k_i\cdot k_j \, P(\omega_{ij})\,,\qquad
\delta  Q = 2 \sum_{1\leq i<j\leq 4} k_i\cdot k_j \delta P(\nu_{ij})\, .
\end{equation}

\subsubsection{Fermionic propagators}
\label{sec:ferprop}

Our normalization for the fermionic propagators in the $(a,b)$-spin
structure is given by

\begin{equation}
  \langle \psi^\mu(z) \psi^\nu(0)\rangle_{one-loop}=
  {\alpha'\over2}\, S_{a,b}(z|\tau)\,.  
\end{equation}

\noindent{$\triangleright$} In the even spin structure fermionic propagators are
\begin{equation}\label{e:eSeven}
S_{a,b}(z|\tau)= {\theta\left[a\atop b\right](z|\tau)\over \theta\left[a\atop b\right](0|\tau)}\, {\partial_z\theta\left[1\atop 1\right](0|\tau)\over \theta\left[1\atop 1\right](z|\tau)}\,.
\end{equation}
The odd spin structure propagator is 
\begin{equation}\label{e:Soddbis}
S_{1,1}(z|\tau)={ \partial_z\theta\left[1\atop 1\right](z|\tau)\over \theta\left[1\atop 1\right](z|\tau)}\,,
\end{equation}
and the fermionic propagator orthogonal to the zero modes is
\begin{equation}\label{e:Soddn}
\tilde S_{1,1}(z|\tau)=S_{1,1}(z|\tau)- 2i\pi{z_2\over \tau_{2}}
=-4\partial_{z}\cP(z|\tau)\,.
\end{equation}
The fermionic propagators have the following $q$ expansion representation
\cite{Gradstein}
\begin{eqnarray}\label{e:SSFourier}
S_{1,1}(z|\tau)&=&{\pi\over \tan(\pi z)}+ 4\pi \sum_{n=1}^{\infty}\, {q^{n}\over 1-q^{n}}\, \sin(2 n \pi z)\,,\cr
S_{1,0}(z|\tau)&=&{\pi\over \tan(\pi z)}- 4\pi \sum_{n=1}^{\infty}\, {q^{n}\over 1+q^{n}}\, \sin(2 n \pi z)\,,\cr
S_{0,0}(z|\tau)&=&{\pi\over \sin(\pi z)}-4\pi \sum_{n=1}^\infty {q^{n-\frac12}\over 1+q^{n-\frac12}}\,\sin((2n-1)\pi z)\,,\cr
S_{0,1}(z|\tau)&=&{\pi\over \sin(\pi z)}+4\pi \sum_{n=1}^\infty {q^{n-\frac12}\over 1-q^{n-\frac12}}\,\sin((2n-1)\pi z)\,.
\end{eqnarray}

\noindent$\triangleright$ Riemann supersymmetric identities written in the text
\eqref{e:Riemann} derive from the following Riemann relation
relation:
\begin{equation}
  \label{e:Jacobi}
  \sum_{a,b=0,1} (-1)^{a+b+ab}\prod_{i=1}^4 \theta\left[a \atop b
\right](v_i)=-2\prod^4_{i=1} \theta_1(v'_i)\,,
\end{equation}
with $v'_1={1\over2} (-v_1+v_2+v_3+v_4)$ $v'_2={1\over2}(v_1-v_2+v_3+v_4)$
$v'_3={1\over2}(v_1+v_2-v_3+v_4)$, and $v'_4={1\over2}(v_1+v_2+v_3-v_4)$.
This identity can be written, in the form used in the main text, as
vanishing identities 
\begin{eqnarray}
  \sum_{a,b=0,1\atop ab=0} (-1)^{a+b+ab} Z_{a,b}(\tau) &=&0\,,\\
  \label{e:Riemann03}\sum_{a,b=0,1\atop ab=0} (-1)^{a+b+ab} Z_{a,b}(\tau) \prod_{r=1}^n
  S_{a,b}(z)&=&0\quad 1\leq n\leq 3\,,
\end{eqnarray}
and the first  non-vanishing one
\begin{equation}
  \label{e:Riemann4}
  \sum_{a,b=0,1\atop ab=0} (-1)^{a+b+ab} Z_{a,b}(\tau) \prod_{i=1}^4 S_{a,b}(z_i|\tau)=-(2\pi)^4\,.
\end{equation}
with $z_1+\cdots+z_4=0$ and where we used that $\partial_z
\theta\left[1\atop1\right](0|\tau)=\pi\theta\left[0\atop0\right](0|\tau)\theta\left[1\atop0\right](0|\tau)\theta\left[0\atop1\right](0|\tau)=2\pi
\eta^3(\tau)$.

Two identities consequences of the Riemann relation in~\eqref{e:Jacobi}
are
\begin{eqnarray}\label{e:IdSsquare}
  S_{0,0}^2(z) - S_{1,0}^2(z)&=&\pi^2 (\theta\left[0\atop
    1\right](0|\tau))^4\,\left(\partial_z\theta\left[1\atop1\right](z|\tau)\over
    \partial\theta\left[1\atop1\right](0|\tau)\right)^2\cr
S_{0,1}^2(z) - S_{1,0}^2(z)&=&\pi^2 (\theta\left[0\atop
    0\right](0|\tau))^4\,\left(\partial_z\theta\left[1\atop1\right](z|\tau)\over
   \partial\theta\left[1\atop1\right](0|\tau)\right)^2\,.
\end{eqnarray}

\subsubsection{$q$ expansion}
\label{sec:q-expansion}

The $q$ expansions of the fermionic propagators in the even spin
structure are given by
\begin{eqnarray}
S_{1,0}(z|\tau)&=&{\pi\over \tan(\pi z)} -4\pi q\sin(2\pi z)+o(q^2)\,,\cr
\label{e:SKq}  S_{0,0}(z|\tau)&=&{\pi\over \sin(\pi z)}-4\pi \sqrt q\,\sin(\pi z)+o(q)\,,\\
\nn S_{0,1}(z|\tau)&=&{\pi\over \sin(\pi z)}+4\pi \sqrt q\,\sin(\pi z)+o(q)\,.
\end{eqnarray}

Setting  $\mathcal S^n_{a,b}= \prod_{i=1}^n  S_{a,b}(z_i|\tau)$ we
have the following expansion
\begin{eqnarray}
  S^n_{1,0}&=&  \prod_{i=1}^n \pi\cot(\pi z_i)\, \left(1 -8q \sum_{i=1}^n \sin^2(\pi z_i)\right)+o(q^2)\,,\nn\\
S^n_{0,0}& =&  \prod_{i=1}^n\pi (\sin(\pi z_i))^{-1}\,
\left(1 -4q \sum_{i=1}^n \sin^2(\pi z_i)\right)+o(q^2)\,,\\
S^n_{0,1}&=&  \prod_{i=1}^n\pi (\sin(\pi z_i))^{-1}\,
\left(1+4q \sum_{i=1}^n \sin^2(\pi z_i)\right)+o(q^2)\,.\nn
\end{eqnarray}
Applying these identities with $n=2$ and $n=4$ we derive the following
relations between the  correlators $\cW^F_{a,b}$ defined
in~\eqref{e:WFabdef}

\begin{equation}
  \cW^F_{0,0}|_{q^0}=\cW^F_{0,1}|_{q^0};\qquad
\cW^F_{0,0}|_{\sqrt q}=-\cW^F_{0,1}|_{\sqrt q}\,.
\end{equation}
Using the $q$ expansion of the bosonic propagator, it is not difficult
to realize that $\cW^B|_{\sqrt q}=0$, and we can promote the
previous relation to the full correlator $\cW_{a,b}$ defined
in~\eqref{e:W1ab} (using   the identities in~\eqref{e:IdSsquare})
\begin{equation}
  \cW_{0,0}|_{q^0}=\cW_{0,1}|_{q^0};\qquad
\cW_{0,0}|_{\sqrt q}=-\cW_{0,1}|_{\sqrt q}\,.
\label{e:Wabrel}\end{equation}
Other useful relations are between the $q$ expansion of the derivative
bosonic propagator $\partial \cP$ and the fermionic propagator
$S_{1,0}$
\begin{eqnarray}
  \label{eq:dPS01}
  \partial_\nu \cP(\nu|\tau)|_{q^0}- {\pi \nu_2\over2i\tau_2}&=& -\frac14
  S_{1,0}(\nu|\tau)|_{q^0}\\
  \nn \partial_\nu \cP(\nu|\tau)|_{q^1}&=&+\frac14
  S_{1,0}(\nu|\tau)|_{q}\,.
\end{eqnarray}
\subsection{Congruence subgroups of $SL(2,\ZZ)$}\label{sec:congruence}
\label{sec:congr-subgr-sl2}
We denote by $SL(2,\ZZ)$ the group of $2\times 2$ matrix with integers
entries of determinant 1. For any $N$ integers we have the following
subgroups of $SL(2,\ZZ)$ 
\begin{eqnarray}
\Gamma_0(N)&=&\left\{\begin{pmatrix}a & b \cr c & d\end{pmatrix} \in
  SL(2,\ZZ) | \begin{pmatrix}a     & b \cr c & d\end{pmatrix} = \begin{pmatrix} * & * \cr
0 & *\end{pmatrix}  \mod N\right\}\,,\nn\\
 \Gamma_1(N)&=&\left\{\begin{pmatrix}a & b \cr c & d\end{pmatrix}  \in SL(2,\ZZ) |
\begin{pmatrix}a & b \cr c & d\end{pmatrix}  =\begin{pmatrix}1 & * \cr
  0 & 1\end{pmatrix} \mod    N\right\}\,,\\
 \Gamma(N)&=&\left\{\begin{pmatrix}a & b \cr c & d\end{pmatrix} \in SL(2,\ZZ) | 
\begin{pmatrix}a   & b \cr c & d\end{pmatrix}  = \begin{pmatrix}1 & 0\cr 0 &
1\end{pmatrix}  \mod N\right\}\,.\nn
  \end{eqnarray}
They satisfy the properties that $\Gamma(N)\subset \Gamma_1(N)\subset
\Gamma_0(N)\subset SL(2,\ZZ)$.
\section{Chiral blocks for the type II orbifolds}\label{sec:chblocks}
We recall some essential facts from the construction of~\cite{Gregori:1997hi}.
The shifted $T^2$ lattice sum writes
\begin{equation}
  \label{e:Gamma22wDef}
  \Gamma_{(2,2)}^w\left[ h\atop g\right] := \sum_{P_L,p_R\in
    \Gamma_{(2,2)}+w {h\over2}} \, e^{i\pi gl\cdot w}
  q^{P_L^2\over2}\bar q^{P^2_R\over2}\,,
\end{equation}
where $\ell\cdot w=m_I b^I + a_I n^I $ where the shift vector
$w=(a_I, b^I)$ is  such that $w^2=2 a\cdot b=0$ and
\begin{eqnarray}
  P_L^2&=& {| U(m_1+ a_1 {h\over2})- (m_2+a_2{h\over2}) +T (n^1+
    b^1{h\over2})+TU (n^2+ b^2{h\over2})|^2\over2T_2U_2}\,,\cr
P^2_L-P^2_R&=&2(m_I+ a_I {h\over2})(n^I+b^I {h\over2})  \,.
\end{eqnarray}
$T$ and $U$ are the moduli of the $T^2$.
We recall the full expressions for the orbifold blocks :

\begin{equation}
\cZ^{(22);h,g}_{a,b} := 
\begin{cases}
  \cZ_{a,b}=\eqref{e:chblkuntw}& (h,g)=(0,0)\cr
 4(-1)^{(a+h)g}\,
\left({\theta\left[a\atop
b\right](0|\tau) \theta\left[a+h\atop b+g\right](0|\tau)\over
\eta(\tau)^3\theta\left[ 1+h \atop 1+g
\right]}\right)^2\times\Gamma_{(2,2)}(T,U)& (h,g)\neq (0,0)\,,
\end{cases}
\label{e:orbsym22}
\end{equation}
\begin{equation}
\cZ^{(14);h,g}_{a,b} = {1\over 2}\sum_{h',g'=0}^1
\cZ^{h,g}_{a,b}\left[h'\atop g'\right]
\Gamma_{(2,2)}^{w}\left[h'\atop g'\right]\,,
\label{e:orbsym14}
\end{equation}
\begin{equation}
\cZ^{(10);h,g}_{a,b} = {1\over 2}\sum_{h_1,g_1=0}^1 {1\over
2}\sum_{h_2,g_2=0}^1
 \cZ^{h,g}_{a,b}\left[h;h_1,h_2\atop g;g_1,g_2\right]
\Gamma_{(2,2)}^{w_1,w_2}\left[h_1,h_2\atop g_1,g_2\right]\, , \, \forall  h,g\, .
\label{e:orbsym10}
\end{equation}

For the $n_v=6$ model, the orbifold acts differently and we get
\begin{equation}
\cZ^{(6);h,g}_{a,b} = {1\over 2}\sum_{h',g'=0}^1
(-1)^ {h g'+gh'}  \cZ^{h,g}_{a,b}
\Gamma_{(2,2)}^w\left[h'\atop g'\right]\,.
\label{e:orbsym6}
\end{equation}

In the previous expressions, the crucial point is that the shifted lattice sums
$\Gamma_{(2,2)}^w\abd {h'}{ g'}$ act as projectors on their untwisted $h'=0$ sector, while
the $g'$ sector is left free. We recall now the diagonal properties of the orbifold action
(see \cite{Gregori:1997hi} again) on the lattice sums:
\begin{equation}
 \Gamma_{(2,2)}^{w_1,w_2}\abd{h,0}{g,0}=
\Gamma_{(2,2)}^{w_1}\abd{h}{g}
,\,
\Gamma_{(2,2)}^{w_1,w_2}\abd{0,h}{0,g}=
\Gamma_{(2,2)}^{w_2}\abd{h}{g}
,\,
\Gamma_{(2,2)}^{w_1,w_2}\abd{h,h}{g,g}=
\Gamma_{(2,2)}^{w_{12}}\abd{h}{g}
\, ,
\label{e:shiftlat22}
\end{equation}

The four-dimensional blocks $ \cZ^{h,g}_{a,b}$ have the following properties :
$\cZ^{h,g}_{a,b}\abd{0}{0}=
\cZ^{h,g}_{a,b}\abd{h}{g}=
\cZ^{h,g}_{a,b}$ (ordinary twist); $\cZ^{0,0}_{a,b}\abd{h}{g}$
is a ($4,4$) lattice sum with one shifted momentum and thus  projects out the $h=0$ sector.
Equivalent properties stand as well for the $n_v=10$ model.

One has then in the field theory limit
\begin{eqnarray}
\cZ^{(14);h,g}_{a,b} &\in&\{ \cZ^{0,0}_{a,b},\,\cZ^{0,1}_{a,b}
,\, \frac12 \cZ^{1,0}_{a,b} ,\, \frac 12 \cZ^{1,1}_{a,b}\} \nn\,,\\
\cZ^{(10);h,g}_{a,b} &\in& \{ \cZ^{0,0}_{a,b},\,\cZ^{0,1}_{a,b}
,\, \frac14 \cZ^{1,0}_{a,b} ,\, \frac 14 \cZ^{1,1}_{a,b}\}
\nn\,,\\
\cZ^{(6);h,g}_{a,b} &\in& \{ \cZ^{0,0}_{a,b},\,\cZ^{0,1}_{a,b}
,\, 0 ,\, 0\}\,,
\label{e:orbsymftl}
\end{eqnarray}
from where we easily deduce the effective definition given in \eqref{e:ZIIsym}
and the number $c_h$.


\begin{thebibliography}{99}


\bibitem{Bern:2012cd}
Z.~Bern, S.~Davies, T.~Dennen and Y.~-t.~Huang,
``Absence of Three-Loop Four-Point Divergences in ${\mathcal{N}}\!=4$ Supergravity,''
Phys.\ Rev.\ Lett.\ {\bf 108} (2012) 201301
[arXiv:1202.3423 [hep-th]].



\bibitem{Tourkine:2012ip}
P.~Tourkine and P.~Vanhove,
``An R$^4$ Non-Renormalisation Theorem in ${\mathcal{N}}\!=4$ Supergravity,''
Class.\ Quant.\ Grav.\ {\bf 29} (2012) 115006
[arXiv:1202.3692 [hep-th]].



\bibitem{Bossard:2011tq}
G.~Bossard, P.~S.~Howe, K.~S.~Stelle and P.~Vanhove,
``The Vanishing Volume of $d=4$ Superspace,''
Class.\ Quant.\ Grav.\ {\bf 28} (2011) 215005
[arXiv:1105.6087 [hep-th]].


\bibitem{Deser:1977nt}
S.~Deser, J.~H.~Kay and K.~S.~Stelle,
``Renormalizability Properties of Supergravity,''
Phys.\ Rev.\ Lett.\ {\bf 38} (1977) 527.




\bibitem{Kallosh:2012ei}
R.~Kallosh,
``On Absence of 3-Loop Divergence in ${\mathcal{N}}\!=4$ Supergravity,''
Phys.\ Rev.\ D {\bf 85} (2012) 081702
[arXiv:1202.4690 [hep-th]].

\bibitem{Bern:2007hh}
Z.~Bern, J.~J.~Carrasco, L.~J.~Dixon, H.~Johansson, D.~A.~Kosower and R.~Roiban,
``Three-Loop Superfiniteness of ${\mathcal{N}}\!=8$ Supergravity,''
Phys.\ Rev.\ Lett.\ {\bf 98} (2007) 161303
[hep-th/0702112].

\bibitem{Bern:2008pv}
Z.~Bern, J.~J.~M.~Carrasco, L.~J.~Dixon, H.~Johansson and R.~Roiban,
``Manifest Ultraviolet Behavior for the Three-Loop Four-Point Amplitude of ${\mathcal{N}}\!=8$ Supergravity,''
Phys.\ Rev.\ D {\bf 78} (2008) 105019
[arXiv:0808.4112 [hep-th]].


\bibitem{Bern:2009kd}
Z.~Bern, J.~J.~Carrasco, L.~J.~Dixon, H.~Johansson and R.~Roiban,
``The Ultraviolet Behavior of ${\mathcal{N}}\!=8$ Supergravity at Four Loops,''
Phys.\ Rev.\ Lett.\ {\bf 103} (2009) 081301
[arXiv:0905.2326 [hep-th]].

\bibitem{Green:2006gt}
M.~B.~Green, J.~G.~Russo and P.~Vanhove,
``Non-Renormalisation Conditions in Type II String Theory and Maximal Supergravity,''
JHEP {\bf 0702} (2007) 099
[hep-th/0610299].

\bibitem{Green:2010sp}
M.~B.~Green, J.~G.~Russo and P.~Vanhove,
``String Theory Dualities and Supergravity Divergences,''
JHEP {\bf 1006} (2010) 075
[arXiv:1002.3805 [hep-th]].

\bibitem{Bossard:2009sy}
G.~Bossard, P.~S.~Howe and K.~S.~Stelle,
``The Ultra-Violet Question in Maximally Supersymmetric Field Theories,''
Gen.\ Rel.\ Grav.\ {\bf 41} (2009) 919
[arXiv:0901.4661 [hep-th]].

\bibitem{Bossard:2010bd}
G.~Bossard, P.~S.~Howe and K.~S.~Stelle,
``On Duality Symmetries of Supergravity Invariants,''
JHEP {\bf 1101} (2011) 020
[arXiv:1009.0743 [hep-th]].

\bibitem{Tseytlin:1995bi}
  A.~A.~Tseytlin,
  ``Heterotic type I superstring duality and low-energy effective actions,''
  Nucl.\ Phys.\ B {\bf 467} (1996) 383
  [hep-th/9512081].

\bibitem{Bachas:1997mc}
C.~Bachas, C.~Fabre, E.~Kiritsis, N.~A.~Obers and P.~Vanhove,
``Heterotic / Type I Duality and D-Brane Instantons,''
Nucl.\ Phys.\ B {\bf 509} (1998) 33
[hep-th/9707126].


\bibitem{D'Hoker:2005jc}
E.~D'Hoker and D.~H.~Phong,
``Two-Loop Superstrings Vi: Non-Renormalization Theorems and the 4-Point Function,''
Nucl.\ Phys.\ B {\bf 715} (2005) 3
[hep-th/0501197].



\bibitem{Elvang:2010jv}
H.~Elvang, D.~Z.~Freedman and M.~Kiermaier,
``A Simple Approach to Counterterms in ${\mathcal{N}}\!=8$ Supergravity,''
JHEP {\bf 1011} (2010) 016
[arXiv:1003.5018 [hep-th]].


\bibitem{Elvang:2010kc}
H.~Elvang and M.~Kiermaier,
``Stringy Klt Relations, Global Symmetries, and E7(7) Violation,''
JHEP {\bf 1010} (2010) 108
[arXiv:1007.4813 [hep-th]].


\bibitem{Beisert:2010jx}
N.~Beisert, H.~Elvang, D.~Z.~Freedman, M.~Kiermaier, A.~Morales and S.~Stieberger,
``E7(7) Constraints on Counterterms in ${\mathcal{N}}\!=8$ Supergravity,''
Phys.\ Lett.\ B {\bf 694} (2010) 265
[arXiv:1009.1643 [hep-th]].




\bibitem{Elvang:2010xn}
H.~Elvang, D.~Z.~Freedman and M.~Kiermaier,
``SUSY Ward Identities, Superamplitudes, and Counterterms,''
J.\ Phys.\ A A {\bf 44} (2011) 454009
[arXiv:1012.3401 [hep-th]].

\bibitem{Green:1982sw}
M.~B.~Green, J.~H.~Schwarz and L.~Brink,
``${\mathcal{N}}\!=4$ Yang-Mills and ${\mathcal{N}}\!=8$ Supergravity as Limits of String Theories,''
Nucl.\ Phys.\ B {\bf 198} (1982) 474.

\bibitem{Bern:1990cu} 
  Z.~Bern and D.~A.~Kosower,
  ``Efficient calculation of one loop QCD amplitudes,''
  Phys.\ Rev.\ Lett.\  {\bf 66}, 1669 (1991).

\bibitem{Bern:1990ux} 
  Z.~Bern and D.~A.~Kosower,
  ``Color decomposition of one loop amplitudes in gauge theories,''
  Nucl.\ Phys.\ B {\bf 362}, 389 (1991).



\bibitem{Bern:1993wt} 
  Z.~Bern, D.~C.~Dunbar and T.~Shimada,
  ``String based methods in perturbative gravity,''
  Phys.\ Lett.\ B {\bf 312}, 277 (1993)
  [hep-th/9307001].



\bibitem{Dunbar:1994bn}
D.~C.~Dunbar and P.~S.~Norridge,
``Calculation of Graviton Scattering Amplitudes Using String Based Methods,''
Nucl.\ Phys.\ B {\bf 433} (1995) 181
[hep-th/9408014].


\bibitem{BjerrumBohr:2008ji}
N.~E.~J.~Bjerrum-Bohr and P.~Vanhove,
``Absence of Triangles in Maximal Supergravity Amplitudes,''
JHEP {\bf 0810} (2008) 006
[arXiv:0805.3682 [hep-th]].



\bibitem{Dabholkar:1998kv}
A.~Dabholkar and J.~A.~Harvey,
``String Islands,''
JHEP {\bf 9902} (1999) 006
[hep-th/9809122].


\bibitem{Witten:1995ex}
E.~Witten,
``String Theory Dynamics in Various Dimensions,''
Nucl.\ Phys.\ B {\bf 443} (1995) 85
[arXiv:hep-th/9503124].

\bibitem{Hull:1994ys}
C.~M.~Hull and P.~K.~Townsend,
``Unity of Superstring Dualities,''
Nucl.\ Phys.\ B {\bf 438} (1995) 109
[arXiv:hep-th/9410167].




\bibitem{Gregori:1997hi}
A.~Gregori, E.~Kiritsis, C.~Kounnas, N.~A.~Obers, P.~M.~Petropoulos and B.~Pioline,
``$ R^2$ Corrections and Non-Perturbative Dualities of ${\mathcal{N}}\!=4$ String Ground States,''
Nucl.\ Phys.\ B {\bf 510} (1998) 423
[arXiv:hep-th/9708062].


\bibitem{Dunbar:1995ed}
D.~C.~Dunbar and P.~S.~Norridge,
``Infinities Within Graviton Scattering Amplitudes,''
Class.\ Quant.\ Grav.\ {\bf 14} (1997) 351
[arXiv:hep-th/9512084].


\bibitem{Dunbar:2010fy}
D.~C.~Dunbar, J.~H.~Ettle and W.~B.~Perkins,
``Perturbative Expansion of $N<8$ Supergravity,''
Phys.\ Rev.\ D {\bf 83} (2011) 065015
[arXiv:1011.5378 [hep-th]].


\bibitem{Dunbar:2011xw}
D.~C.~Dunbar, J.~H.~Ettle and W.~B.~Perkins,
``Obtaining One-Loop Gravity Amplitudes Using Spurious Singularities,''
Phys.\ Rev.\ D {\bf 84} (2011) 125029
[arXiv:1109.4827 [hep-th]].

\bibitem{Dunbar:2011dw}
D.~C.~Dunbar, J.~H.~Ettle and W.~B.~Perkins,
``The N-Point Mhv One-Loop Amplitude in ${\mathcal{N}}\!=4$ Supergravity,''
Phys.\ Rev.\ Lett.\ {\bf 108} (2012) 061603
[arXiv:1111.1153 [hep-th]].



\bibitem{Dunbar:2012aj}
D.~C.~Dunbar, J.~H.~Ettle and W.~B.~Perkins,
``Constructing Gravity Amplitudes from Real Soft and Collinear Factorisation,''
Phys.\ Rev.\ D {\bf 86} (2012) 026009
[arXiv:1203.0198 [hep-th]].



\bibitem{Bern:2011rj}
Z.~Bern, C.~Boucher-Veronneau and H.~Johansson,
``$N\geq 4$ Supergravity Amplitudes from Gauge Theory at One Loop,''
Phys.\ Rev.\ D {\bf 84} (2011) 105035
[arXiv:1107.1935 [hep-th]].



\bibitem{Ferrara:1989nm}
S.~Ferrara and C.~Kounnas,
``Extended Supersymmetry in Four-Dimensional Type Ii Strings,''
Nucl.\ Phys.\ B {\bf 328} (1989) 406.


\bibitem{Chaudhuri:1995bf}
S.~Chaudhuri and J.~Polchinski,
``Moduli Space of Chl Strings,''
Phys.\ Rev.\ D {\bf 52} (1995) 7168
[arXiv:hep-th/9506048].



\bibitem{Schwarz:1995bj}
J.~H.~Schwarz and A.~Sen,
``Type Iia Dual of the Six-Dimensional Chl Compactification,''
Phys.\ Lett.\ B {\bf 357} (1995) 323
[arXiv:hep-th/9507027].



\bibitem{Aspinwall:1995fw}
P.~S.~Aspinwall,
``Some Relationships Between Dualities in String Theory,''
Nucl.\ Phys.\ Proc.\ Suppl.\ {\bf 46} (1996) 30
[arXiv:hep-th/9508154].


\bibitem{Green:1987mn}
M.~B.~Green, J.~H.~Schwarz and E.~Witten,
``Superstring Theory. Vol. 2: Loop Amplitudes, Anomalies and Phenomenology,''
(1987)  ( Cambridge Monographs On Mathematical Physics)


\bibitem{Sakai:1986bi}
N.~Sakai and Y.~Tanii,
``One Loop Amplitudes and Effective Action in Superstring Theories,''
Nucl.\ Phys.\ B {\bf 287} (1987) 457.

\bibitem{Narain:1986am}
K.~S.~Narain, M.~H.~Sarmadi and E.~Witten,
``A Note on Toroidal Compactification of Heterotic String Theory,''
Nucl.\ Phys.\ B {\bf 279} (1987) 369.


\bibitem{Jatkar:2005bh}
D.~P.~Jatkar and A.~Sen,
``Dyon Spectrum in Chl Models,''
JHEP {\bf 0604} (2006) 018
[arXiv:hep-th/0510147].



\bibitem{Dabholkar:2006bj}
A.~Dabholkar and D.~Gaiotto,
``Spectrum of Chl Dyons from Genus-Two Partition Function,''
JHEP {\bf 0712} (2007) 087
[arXiv:hep-th/0612011].


\bibitem{Mizoguchi:2001cp}
S.~'y.~Mizoguchi,
``On Asymmetric Orbifolds and the D = 5 No-Modulus Supergravity,''
Phys.\ Lett.\ B {\bf 523} (2001) 351
[hep-th/0109193].

\bibitem{Sen:1995ff}
A.~Sen and C.~Vafa,
``Dual Pairs of Type II String Compactification,''
Nucl.\ Phys.\ B {\bf 455} (1995) 165
[arXiv:hep-th/9508064].

\bibitem{Bern:1991aq}
Z.~Bern and D.~A.~Kosower,
``The Computation of Loop Amplitudes in Gauge Theories,''
Nucl.\ Phys.\ B {\bf 379} (1992) 451.


\bibitem{Green:1999pv}
M.~B.~Green and P.~Vanhove,
``The Low Energy Expansion of the One-Loop Type II Superstring Amplitude,''
Phys.\ Rev.\ D {\bf 61} (2000) 104011
[arXiv:hep-th/9910056].


\bibitem{Bern:2007xj}
Z.~Bern, J.~J.~Carrasco, D.~Forde, H.~Ita and H.~Johansson,
``Unexpected Cancellations in Gravity Theories,''
Phys.\ Rev.\ D {\bf 77} (2008) 025010
[arXiv:0707.1035 [hep-th]].




\bibitem{de Roo:1984gd}
M.~de Roo,
``Matter Coupling in ${\mathcal{N}}\!=4$ Supergravity,''
Nucl.\ Phys.\ B {\bf 255} (1985) 515.

\bibitem{'tHooft:1974bx}
G.~'t Hooft and M.~J.~G.~Veltman,
``One Loop Divergencies in the Theory of Gravitation,''
Annales Poincare Phys.\ Theor.\ A {\bf 20} (1974) 69.



\bibitem{Grisaru:1981ye}
M.~T.~Grisaru and W.~Siegel,
``The One Loop Four Particle S Matrix in Extended Supergravity,''
Phys.\ Lett.\ B {\bf 110} (1982) 49.



\bibitem{Fischler:1979yk}
M.~Fischler,
``Finiteness Calculations for $O(4)$ Through $O(8)$ Extended Supergravity and $O(4)$ Supergravity Coupled to Selfdual $O(4)$ Matter,''
Phys.\ Rev.\ D {\bf 20} (1979) 396.

\bibitem{Gradstein}
I.S.~Gradshteyn, and I.W~Ryzhik, {\sl Table of Integrals, Series, and
  Products} (New York: Academic Press), 1980

\end{thebibliography}
\end{document}